\documentclass[12pt]{article}
\usepackage{graphicx}
\usepackage{amsmath,amssymb}
\usepackage{setspace}

\newcommand{\be}{\begin{equation}}
\newcommand{\ee}{\end{equation}}
\newcommand{\tr}{{\rm Tr}}
\newcommand{\rt}{{\rm tr}}
\newcommand{\Wg}{{\rm W}}
\newcommand{\U}{{\mathcal{U}}}
\newcommand{\OO}{{\mathcal{O}}}
\newcommand{\M}{{\mathcal{M}}}

\begin{document}

\title{Random stochastic matrices from classical compact Lie groups and symmetric spaces}
\author{Lucas H. Oliveira and Marcel Novaes\\Instituto de F\'isica, Universidade Federal de Uberl\^andia\\ Uberl\^andia, MG, 38408-100, Brazil}
\date{}

\maketitle
\begin{abstract}
We consider random stochastic matrices $M$ with elements given by $M_{ij}=|U_{ij}|^2$, with $U$ being uniformly distributed on one of the classical compact Lie groups or some of the associated symmetric spaces. We observe numerically that, for large dimensions, the spectral statistics of $M$, discarding the Perron-Frobenius eigenvalue $1$, are similar to those of the Gaussian Orthogonal ensemble for symmetric matrices and to those of the real Ginibre ensemble for non-symmetric matrices. We compute some spectral statistics using Weingarten functions and establish connections with some difficult enumerative problems involving permutations.

\end{abstract}

\onehalfspacing
\section{Introduction}

Markov chains are a fundamental statistical model with numerous applications, that range from computation and physics to chemistry and biology. A Markov chain in discrete time taking place on $N$ nodes is described by a $N\times N$ stochastic matrix $M$. The elements of such a matrix are probabilities; they are real, satisfy $0\le M_{ij}\le 1$ and the columns are normalized as $\sum_{i=1}^NM_{ij}=1$. Matrices that are bistochastic have the rows normalized as the columns, $\sum_{j=1}^NM_{ij}=1$. 

Every stochastic matrix has $1$ as an eigenvalue, by virtue of the Perron-Frobenius theorem, and we call this the `Perron-Frobenius eigenvalue'. The rest of the spectrum we call the `reduced spectrum'.

Ensembles of random stochastic matrices have been defined imposing independent columns with Dirichlet distributions \cite{horvat,chafai,bordenave,inocentini}. It was initially observed and later proved that, for large dimensions, the reduced spectrum becomes approximately uniformly distributed in a disk in the complex plane, with a small concentration of order $N^{1/2}$ eigenvalues on the real line. The radius of the reduced spectrum decays like $N^{-1/2}$. Moreover, the singular spectrum has a quarter-circle distribution. These results are in line with a general expectation from random matrix theory that real non-symmetric matrices should behave like the real Ginibre ensemble, which has precisely those properties \cite{circ1,circ2,edelman}.

Ensembles of random bistochastic matrices have also been investigated \cite{diac1,bi}. A particular important type of bistochastic matrices are the so-called unistochastic matrices, whose elements are given by $M_{ij}=|U_{ij}|^2$ for some unitary matrix $U$. They arise, for example, in the context of quantum graphs \cite{uzy,tanner1,tanner2}. Berkolaiko has shown \cite{greg} that also in this case the modulus of the second-largest eigenvalue decays like $N^{-1/2}$. \.{Z}yczkowski \emph{et.al.} investigated \cite{zycz} unistochastic and also orthostochastic matrices (whose elements are given by $M_{ij}=O_{ij}^2$ for some orthogonal matrix $O$) at low dimensions and found very interesting spectral results in terms of hypocycloids. 

We generalize this connection between stochastic matrices and Lie group theory by taking into account the remaining classical compact group, the unitary symplectic group, and defining symplectostochastic matrices. Moreover, we also consider random stochastic matrices induced from some compact symmetric spaces, namely the so-called circular ensembles (Cartan classes AI and AII) and the chiral ensembles (classes AII, BDI and CII). All these spaces have natural probability measures, induced from Haar measure; they are discussed in Section 2, along with the corresponding stochastic matrices. 

We refer to all such ensembles of stochastic matrices as Lie-stochastic ensembles. We denote by $\Sigma_O$, $\Sigma_U$ and $\Sigma_S$ those associated with the orthogonal, unitary and symplectic groups, respectively, by $\Sigma_{AI}$ and $\Sigma_{AII}$ those related to the circular ensembles and by $\Sigma_{AIII}$, $\Sigma_{BDI}$ and $\Sigma_{CII}$ those related to the chiral ensembles (results for the other compact two classes of symmetric spaces, $CI$ and $DIII$, are very similar to $CII$ and $BDI$, respectively; we omit their discussion for simplicity).

Although the matrix elements in our Lie-stochastic matrices are not independent, the constraint of stochasticity becomes weak for large dimensions and we may expect universal results in this regime. We find numerically -- and verify algebraically to some extent -- that this indeed seems to be the case. $\Sigma_O$, $\Sigma_U$ and $\Sigma_S$, for instance, seem to have real Ginibre statistics. On the other hand, the ensembles related to the symmetric spaces are, to our knowledge, the first ensembles to be defined that contain bistochastic matrices that are also symmetric. As such, they have real eigenvalues. Universality theory predicts spectral statistics similar to that of the Gaussian Orthogonal ensemble, such as a density of eigenvalues given by the famous Wigner semicircle \cite{wigner0,wigner1,wigner2}. This is what we observe, except that $\Sigma_{AIII}$, $\Sigma_{BDI}$ and $\Sigma_{CII}$ have a free parameter $\alpha$, and satisfy universality only for $\alpha=0$. Our numerical simulations are shown in Section 3. 

An algebraic approach to proving the observed universality can be developed by writing quantities like $\langle\tr M^n\rangle$ and $\langle\tr (MM^T)^n\rangle$ in terms of matrix elements and using the machinery of Weingarten functions, which are known for the groups and symmetric spaces we consider \cite{collins,CS,Banica,matsumoto}. After some background material in Section 4, this approach is discussed in Sections 5, 6 and 7. We obtain partial results that are consistent with universality. However, are not able to carry this program to completion because it leads to some difficult combinatorial problems involving permutation groups.

In Section 8, we review these combinatorial problems. We draw attention to them because they may be of interest in themselves. Large $N$ asymptotics is controlled by another class of combinatorial problems, involving factorizations of permutations. Universality rests in the interplay of those two classes of combinatorial problems. This is discussed in Section 9. We present our conclusions in Section 10.

\section{Lie-stochastic ensembles}

All the classical compact Lie groups and symmetric spaces consist of unitary matrices. As we have see, given such a matrix $U$ it follows trivially that the matrix $M$ with elements given by $M_{ij}=|U_{ij}|^2$ is stochastic (actually, bistochastic). 

If we impose no further condition on $N\times N$ unitary matrices, we have the unitary group $\U(N)$. If we impose that $U$ is real, we have the orthogonal group $\OO(N)$. We write $\Sigma_U$ and $\Sigma_O$ to represent the ensembles of the corresponding unistochastic and orthostochastic matrices, respectively.

Let $S$ be a complex unitary $2N\times 2N$ matrix and $S^D=JS^TJ^T$ its dual matrix, where $J=\left( \begin{array}{cc} 0_N & I_N \\ -I_N & 0_N
\end{array}  \right)$ and $0_N$ and $I_N$ are the $N\times N$ zero and identity matrices. The set of all $S$ satisfying $SS^D=1$ (which are quaternion matrices) is the unitary symplectic group $Sp(2N)$. We write $\Sigma_S$ to represent the ensemble of associated symplectostochastic matrices.

The symmetric spaces $\U(N)/\OO(N)$ and $\U(2N)/Sp(2N)$ are known as the circular ensembles in random matrix theory and denoted AI and AII in the Cartan classification. They can be represented, respectively, by symmetric matrices given by $U=VV^T$ or self-dual matrices given by $U=VV^D$, with $V\in\U(N)$, but we actually choose to represent AII with matrices of the form $U=VV^DJ=VJV^T$. The corresponding ensembles of stochastic matrices we denote by $\Sigma_{AI}$ and $\Sigma_{AII}$. Notice that, in contrast to $\Sigma_U$, $\Sigma_O$ and $\Sigma_S$, the matrices from $\Sigma_{AI}$ and $\Sigma_{AII}$ are symmetric.

The symmetric spaces $G(N)/G(a)\times G(b)$, with $a+b=N$, are known as the chiral ensembles in random matrix theory and are denoted AIII, BDI and CII in the Cartan classification, when $G=\U(N)$, $\OO(N)$ and $Sp(2N)$, respectively. Let $\tilde{J}_a=I_a\oplus(-I_{b})$ and let $\tilde{K}_a=\tilde{J}_a\oplus\tilde{J}_a$. Then elements from AIII and BDI can be represented by matrices given by $U=V\tilde{J}_{a}V^\dag$ and by $U=V\tilde{J}_{a}V^T$, respectively, while elements from CII can be represented by matrices given by $U=V\tilde{K}_{a}V^D$. We write $\Sigma_{AIII}$, $\Sigma_{BDI}$ and $\Sigma_{CII}$ to denote the corresponding ensembles of symmetric stochastic matrices.

\begin{figure}[t]
\includegraphics[scale=0.25,clip]{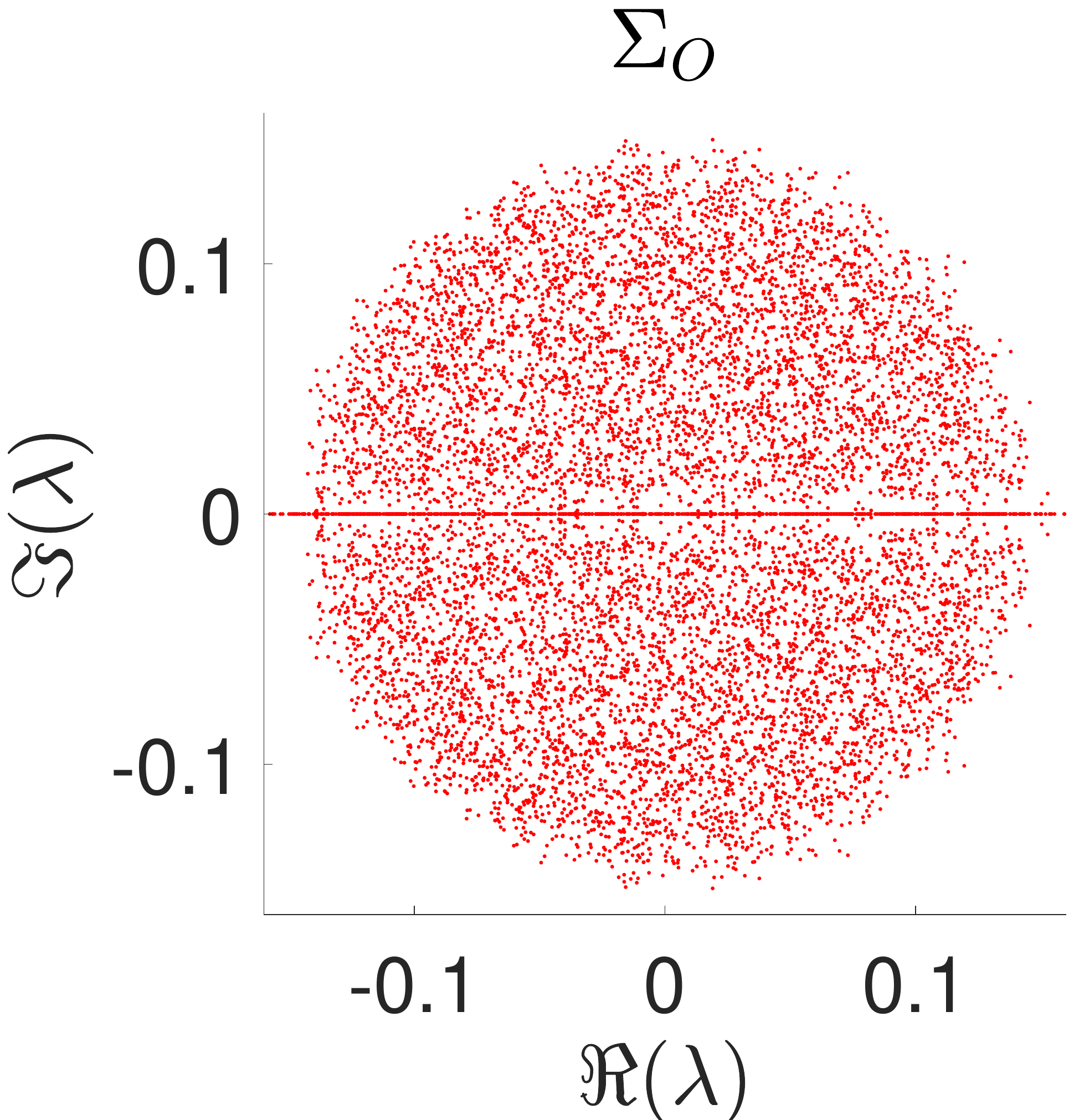}
\includegraphics[scale=0.25,clip]{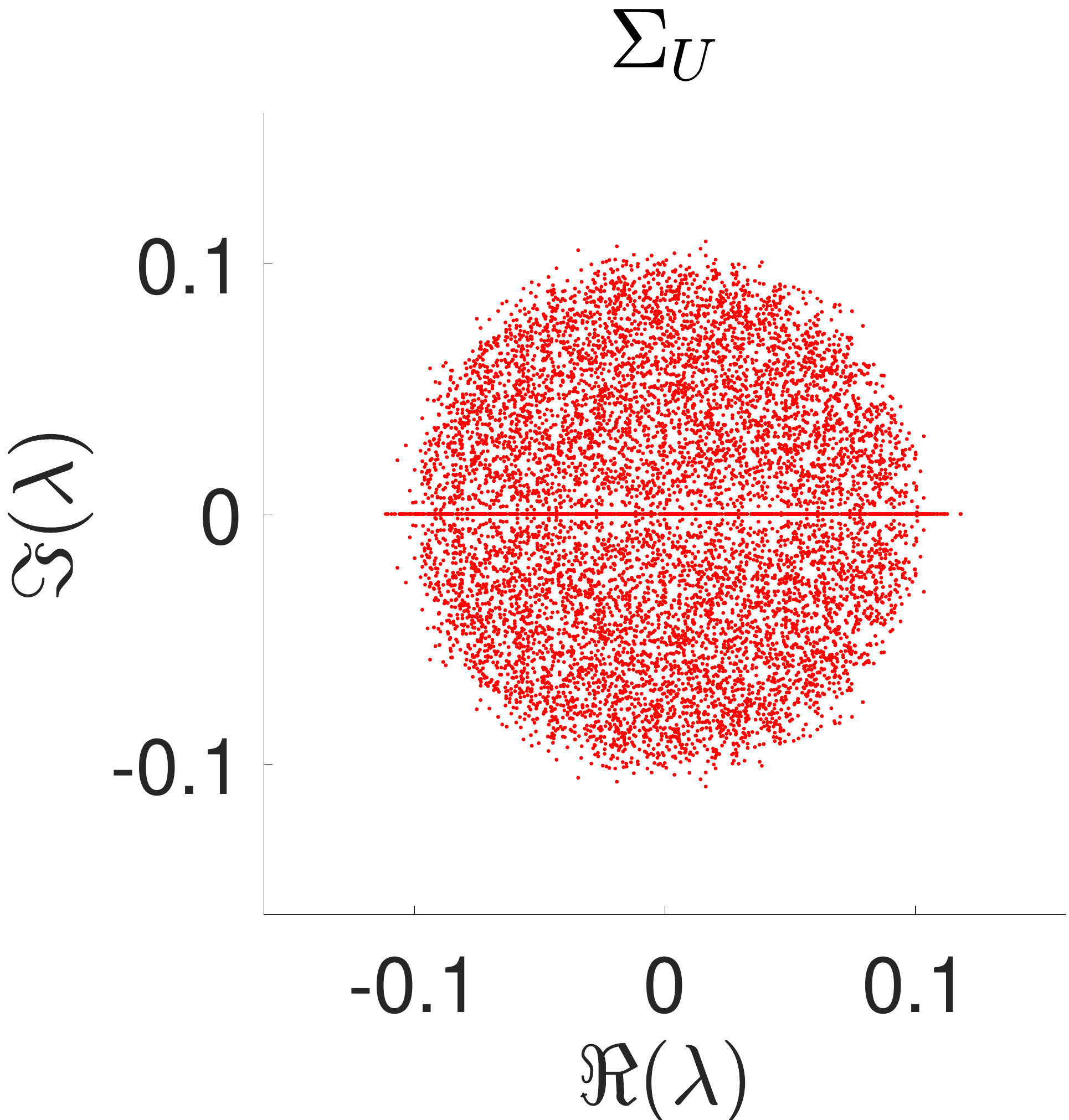}\includegraphics[scale=0.25,clip]{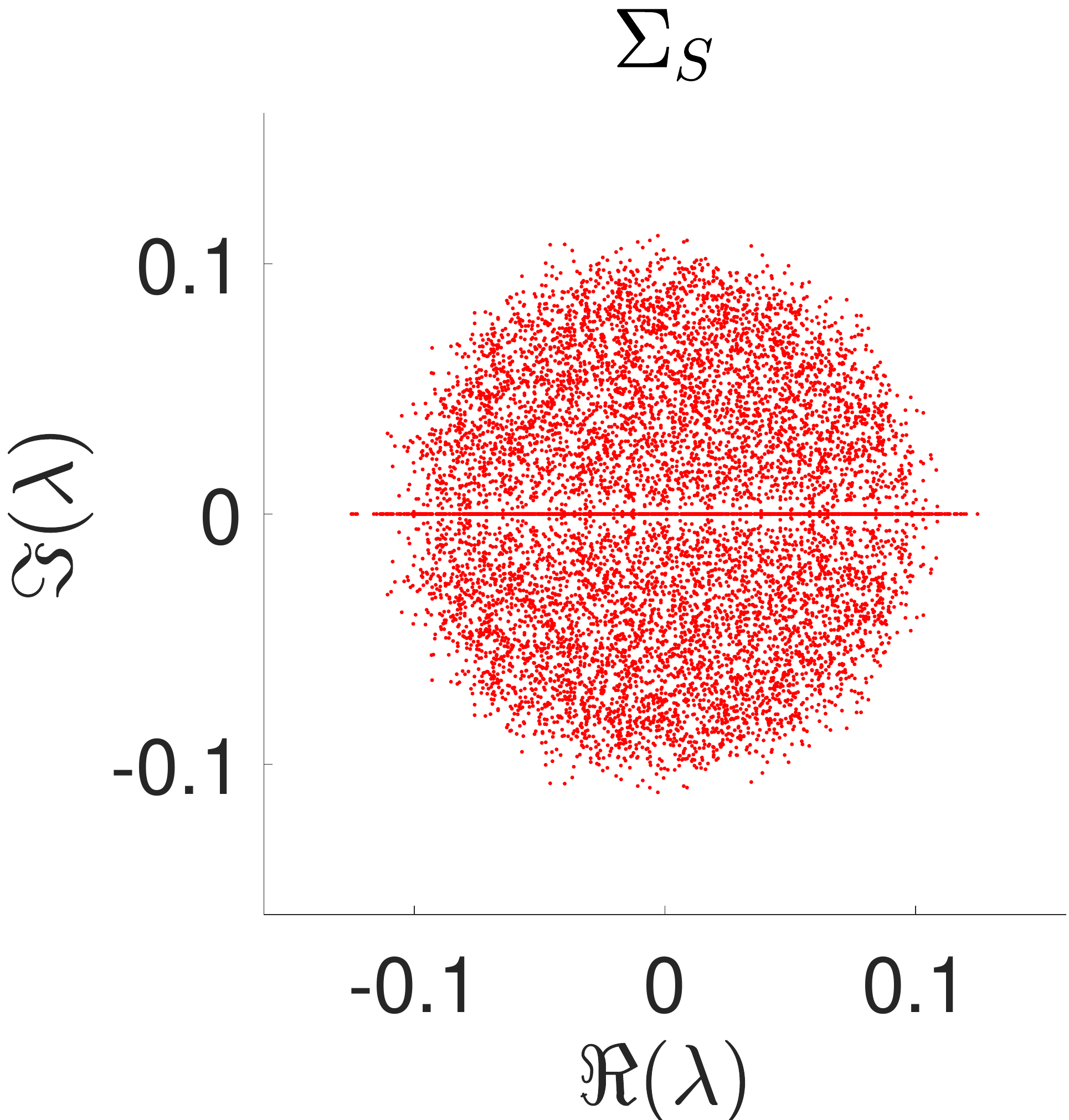}
\caption{Spectrum of 100 realizations of $100\times 100$ random matrices from $\Sigma_O$, $\Sigma_U$ and $\Sigma_S$, respectively. Distribution is similar to that of the real Ginibre ensemble.} \label{Fig1}
\end{figure}
\begin{figure}[t]
\includegraphics[scale=0.27,clip]{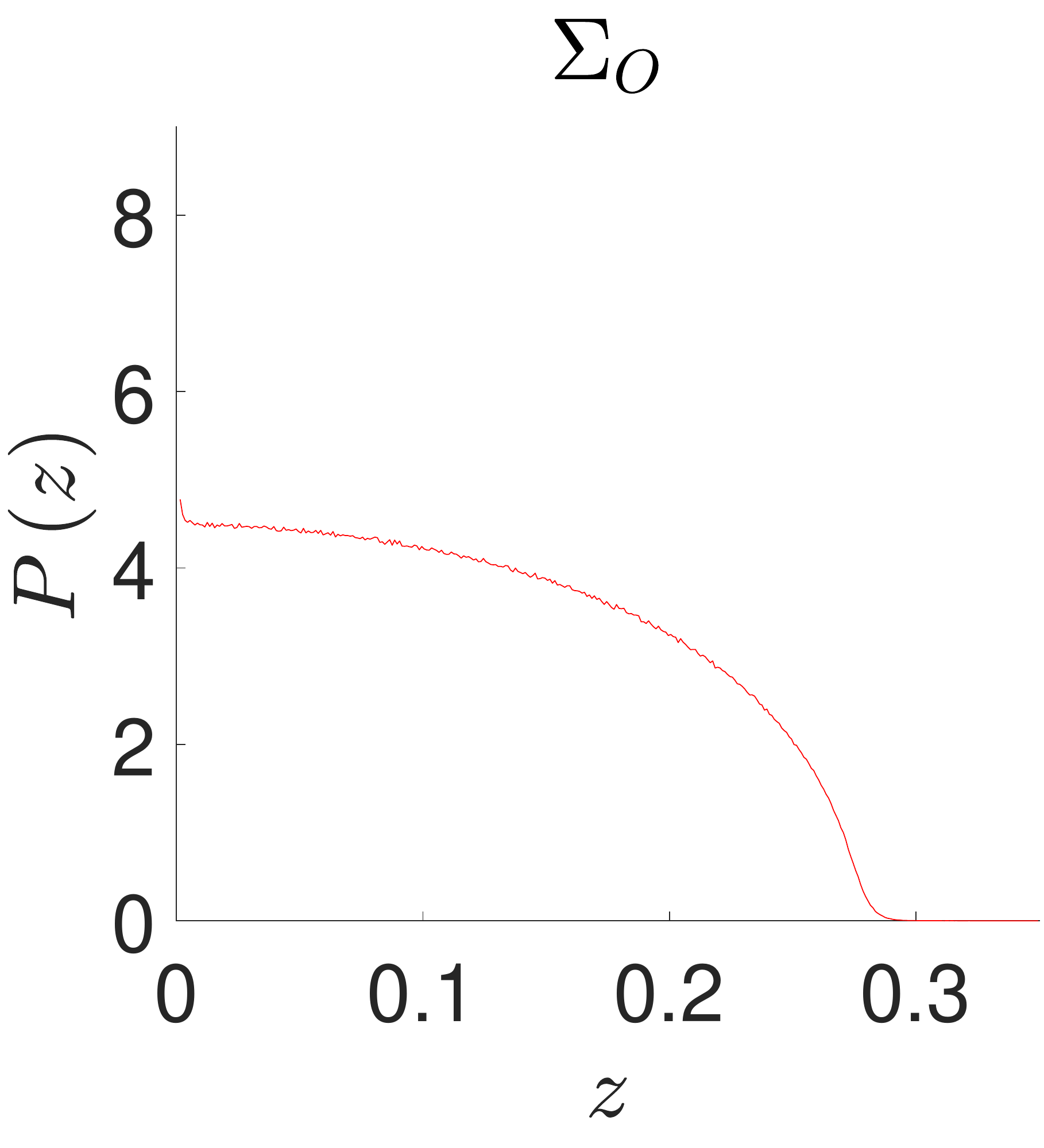}\includegraphics[scale=0.27,clip]{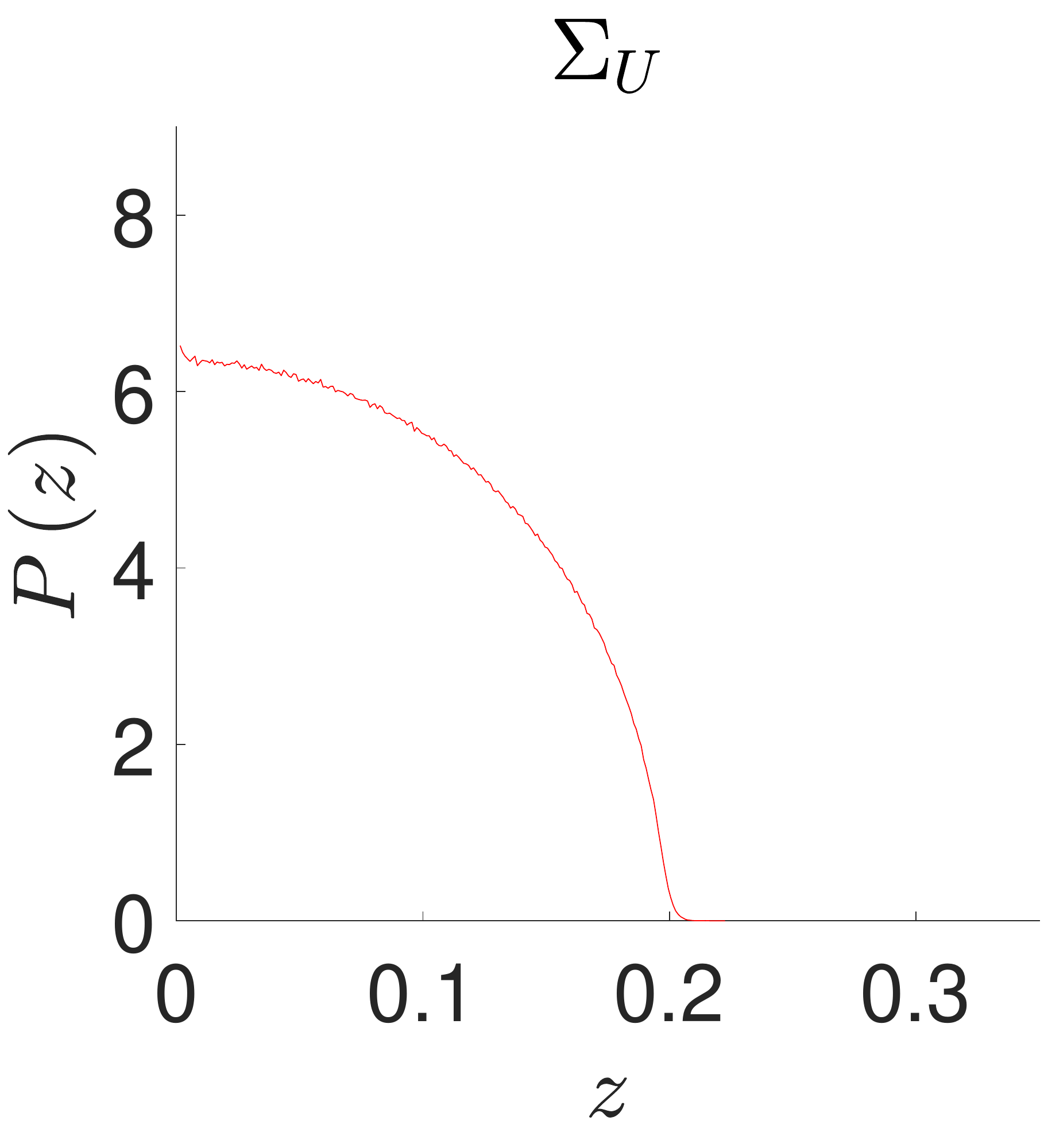}\includegraphics[scale=0.27,clip]{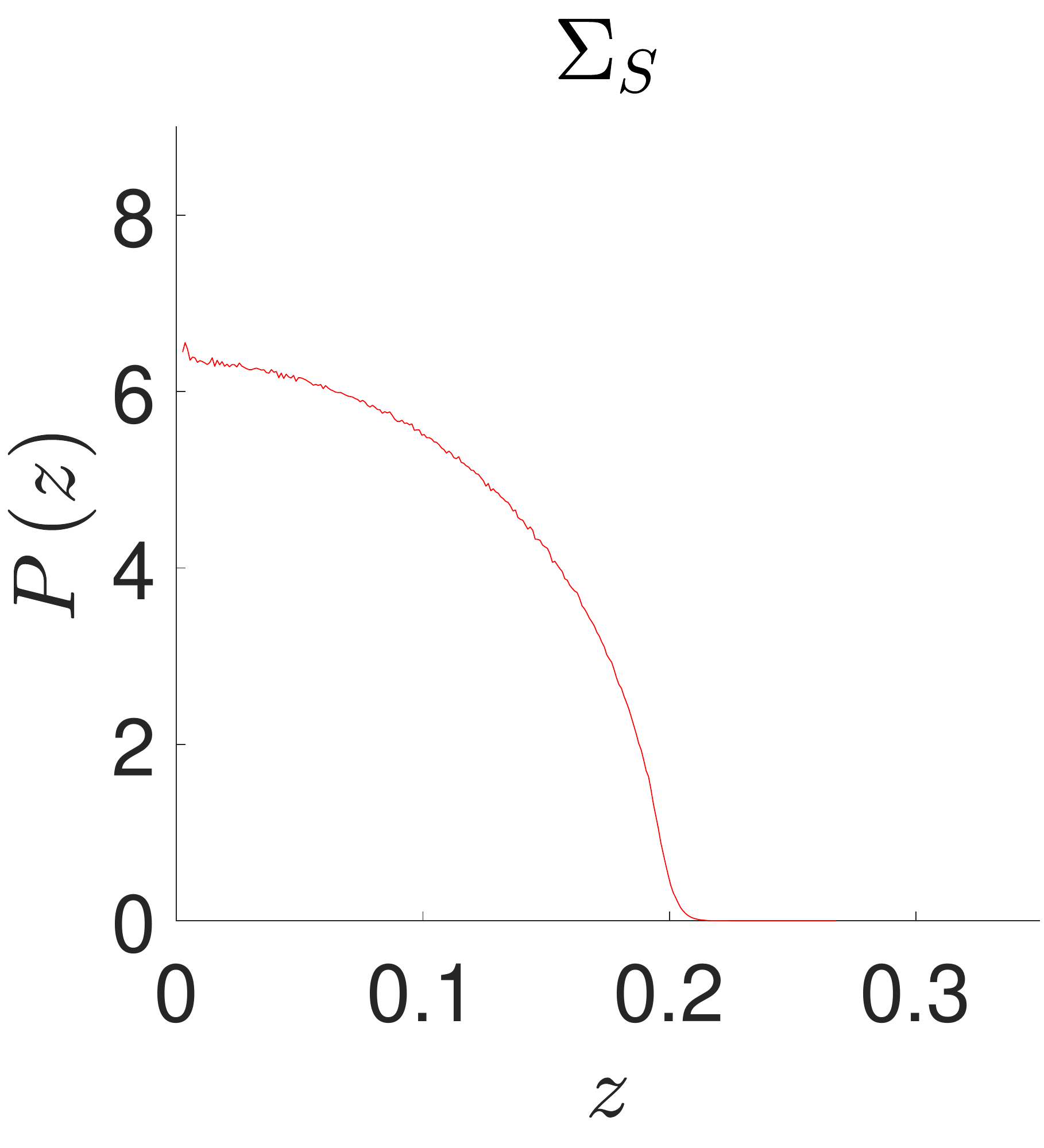}
\caption{Distribution of singular values of $100\times 100$ random matrices from $\Sigma_O$, $\Sigma_U$ and $\Sigma_S$, respectively. Well described by a quarter-circle law.} \label{Fig2}
\end{figure}  

Notice that the extreme cases $a=0$ and $a=N$ are trivial, as $\tilde{J}_a$ becomes proportional to the identity in both cases and so does $U$. When considering the chiral ensembles with large $N$, we adopt a fixed asymmetry parameter $\alpha =(a-b)/N$, taken to be of order 1 and ranging from $-1$ to $1$. Spectral statistics have the obvious symmetry $\alpha\mapsto -\alpha$. 

\section{Numerical results}

We are interested in the regime of large dimensions, $N\gg 1$. Random matrices from Lie groups, distributed uniformly with respect to Haar measure, are generated using QR decomposition, as explained in \cite{mezzadri}. 

Every stochastic matrix has 1 as eigenvalue. In our ensembles of random Lie-stochastic matrices, this eigenvalue is non-degenerate with probability one. Let us define the \emph{reduced spectrum} of a stochastic matrix to mean the set of all its eigenvalues except the 1. Accordingly, let $\rt M$ denote the \emph{reduced trace} of $M$, which is the sum over the reduced spectrum of $M$:
\be \rt M=\sum_{1\neq \lambda\in{\rm spec}(M)}\lambda.\ee

We show in Figure 1 the reduced spectrum of many matrices from $\Sigma_O$, $\Sigma_U$ and $\Sigma_S$ of dimension $N=100$ ($2N=100$ in the symplectic case). These matrices are not symmetric and the distribution of eigenvalues is very similar to that of the real Ginibre ensemble. The points are approximately uniformly distributed inside a disc, except for a small concentration of points in the real line (which seems to contain $\sim\sqrt{N}$ points). The radius of the disc is approximately $\sqrt{2/N}$, $1/\sqrt{N}$ and $1/\sqrt{2N}$, respectively (these can be written as $\sqrt{\frac{2}{\beta N}}$ where $\beta$ is the Dyson index, equal to $1$, $2$, $4$ for $\OO$, $\U$ and $Sp$, respectively). In fact, it seems all eigenvalues in the reduced spectrum are of order $N^{-1/2}$, which is consistent with the calculations we present in the next Section. 

\begin{figure}[t]
\includegraphics[scale=0.3,clip]{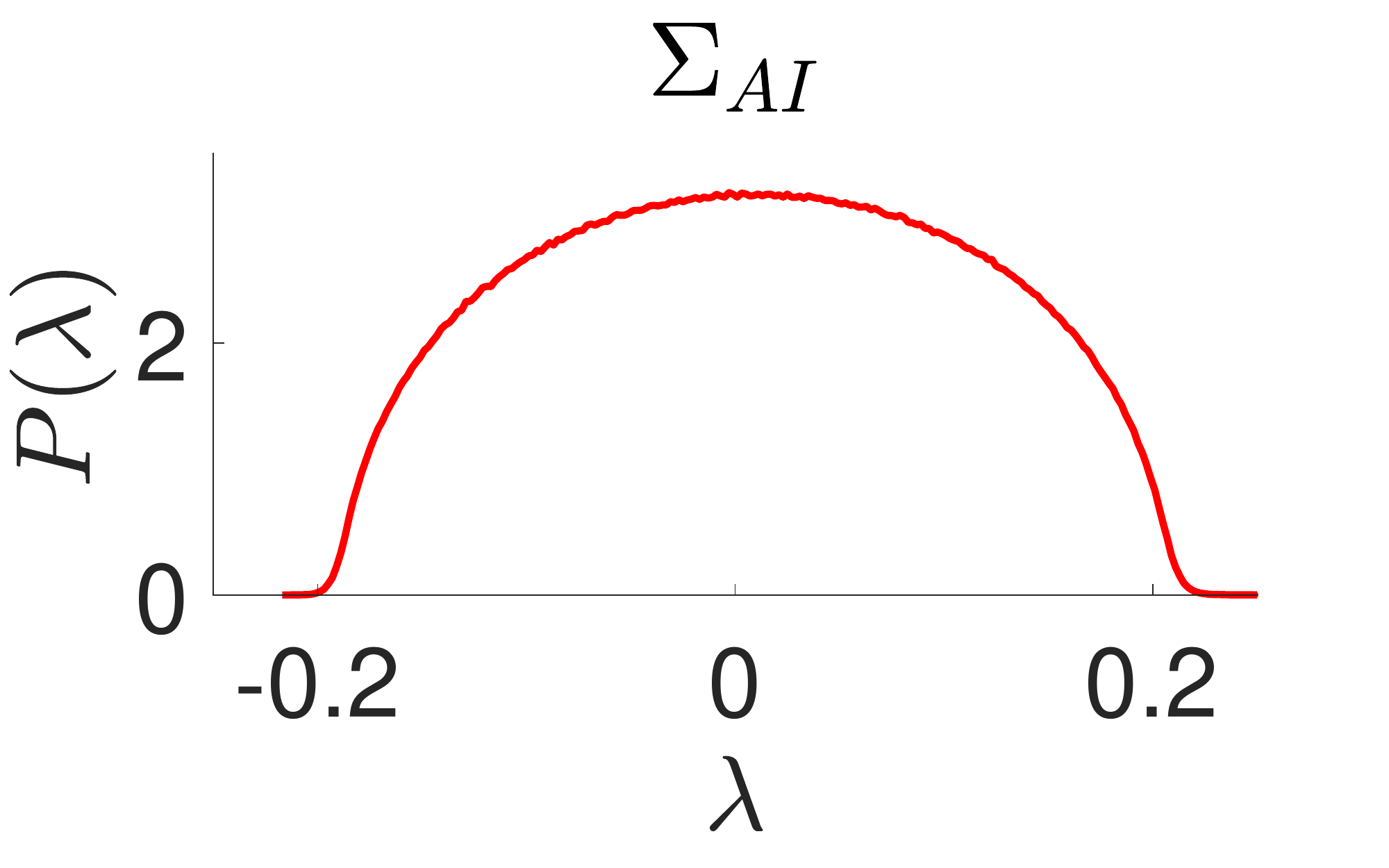}\includegraphics[scale=0.3,clip]{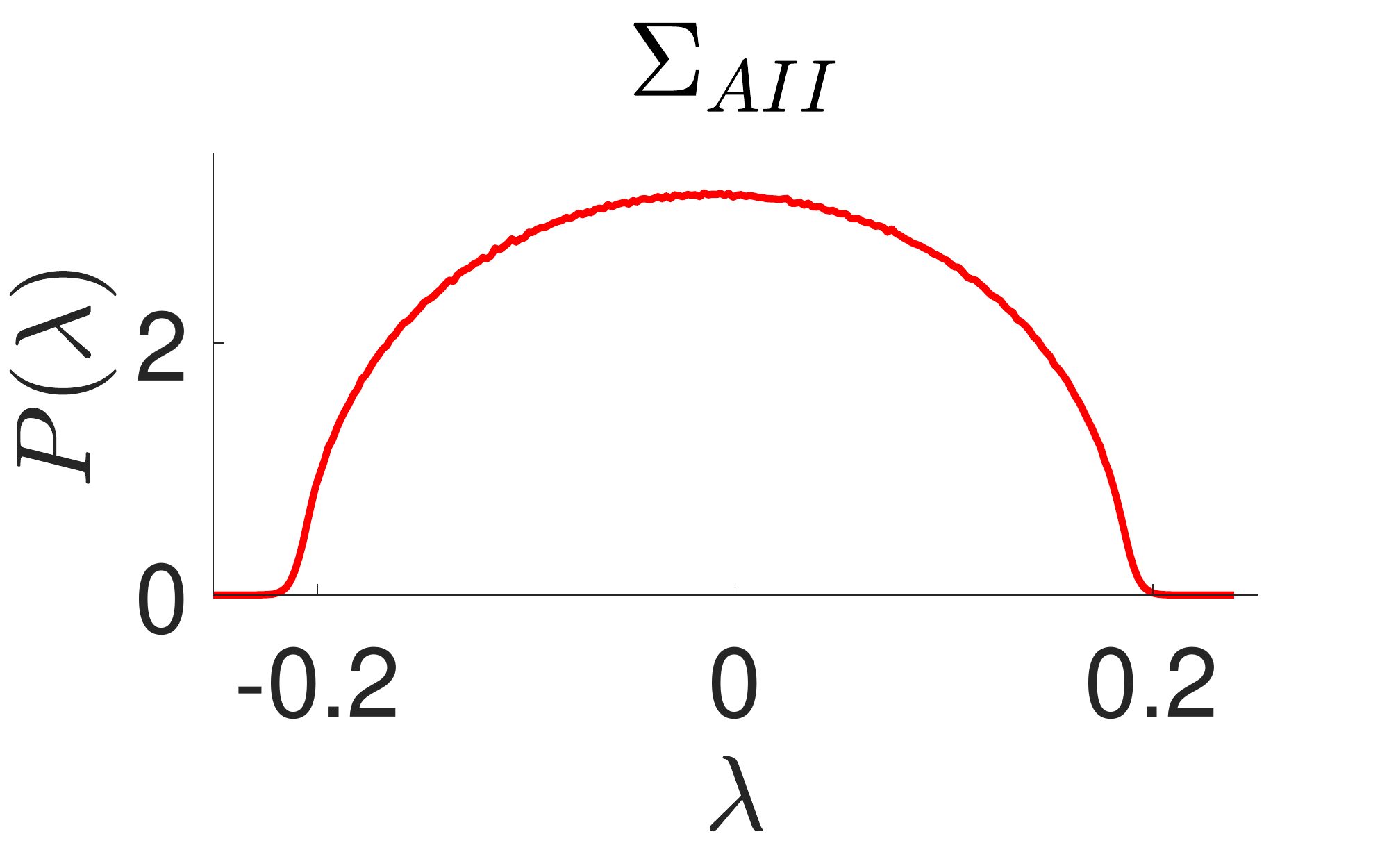}
\caption{Distribution of eigenvalues of $100\times 100$ random matrices from $\Sigma_{AI}$ and $\Sigma_{AII}$. Well described by semi-circle laws.} \label{Fig3}
\end{figure}

The histograms for the reduced singular spectra are shown in Figure 2, again for $N=100$ ($2N=100$ in the symplectic case). The singular values are well described by a quarter-circle distribution, which is consistent with universality. The largest singular value is roughly twice the modulus of the largest eigenvalue, as happens for the real Ginibre ensemble.

In Figure 3 we show the histograms for the real reduced spectra of matrices from $\Sigma_{AI}$ and $\Sigma_{AII}$. They are very well described by a semi-circle law of radius $2/\sqrt{N}$. Such scale is expected since matrices from the usual GOE have elements of order $1$ and a semi-circle of radius $\sqrt{N}$, while our matrices have elements of order $1/N$. However, this universal result is not trivial since, in our ensembles, matrix elements are not independent as required by current universality proofs.

Finally, in Figure 4 we show histograms for the real reduced spectra of matrices from $\Sigma_{BDI}$,  $\Sigma_{AIII}$ and $\Sigma_{CII}$, for three different values of the parameter $\alpha$. For $\alpha=0$ they seem to be semi-circles, but for general $\alpha$ some new distributions appear (we have not attempted to find a numerical fit). As discussed in the previous Section, all eigenvalues converge to $1$ as $\alpha\to 1$.

\begin{figure}[t]
\includegraphics[scale=0.3,clip]{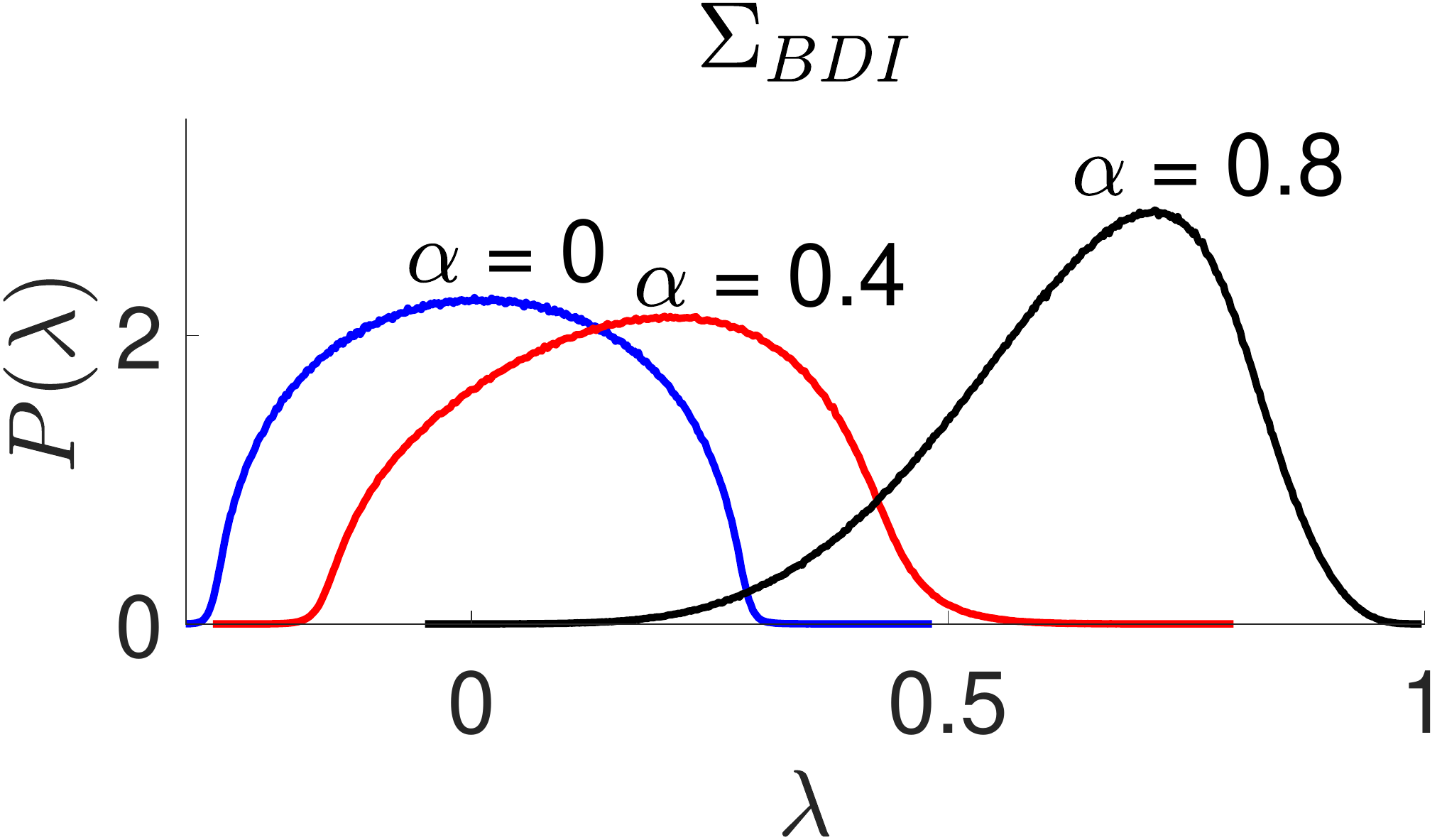}\includegraphics[scale=0.3,clip]{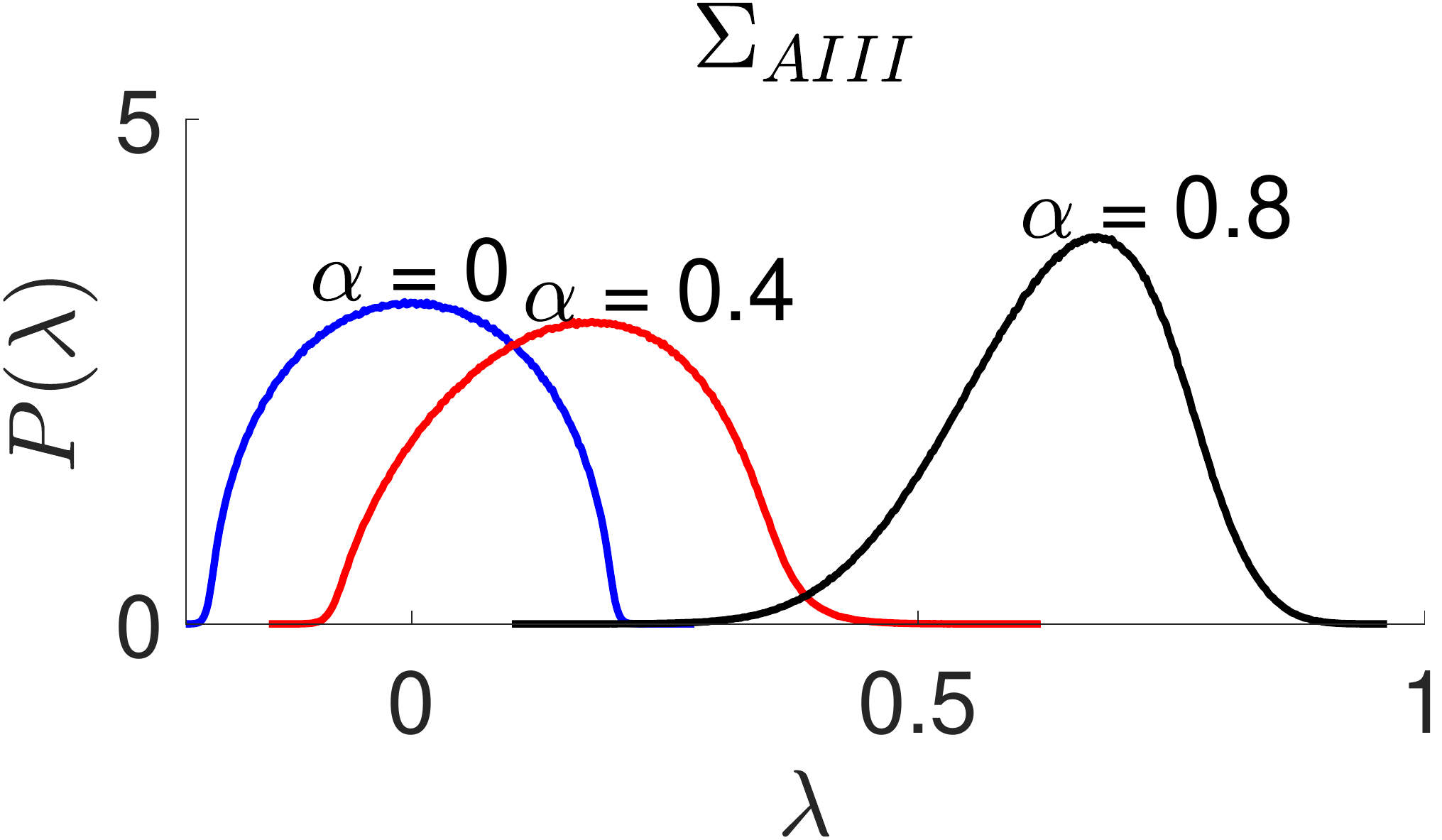}
\includegraphics[scale=0.3,clip]{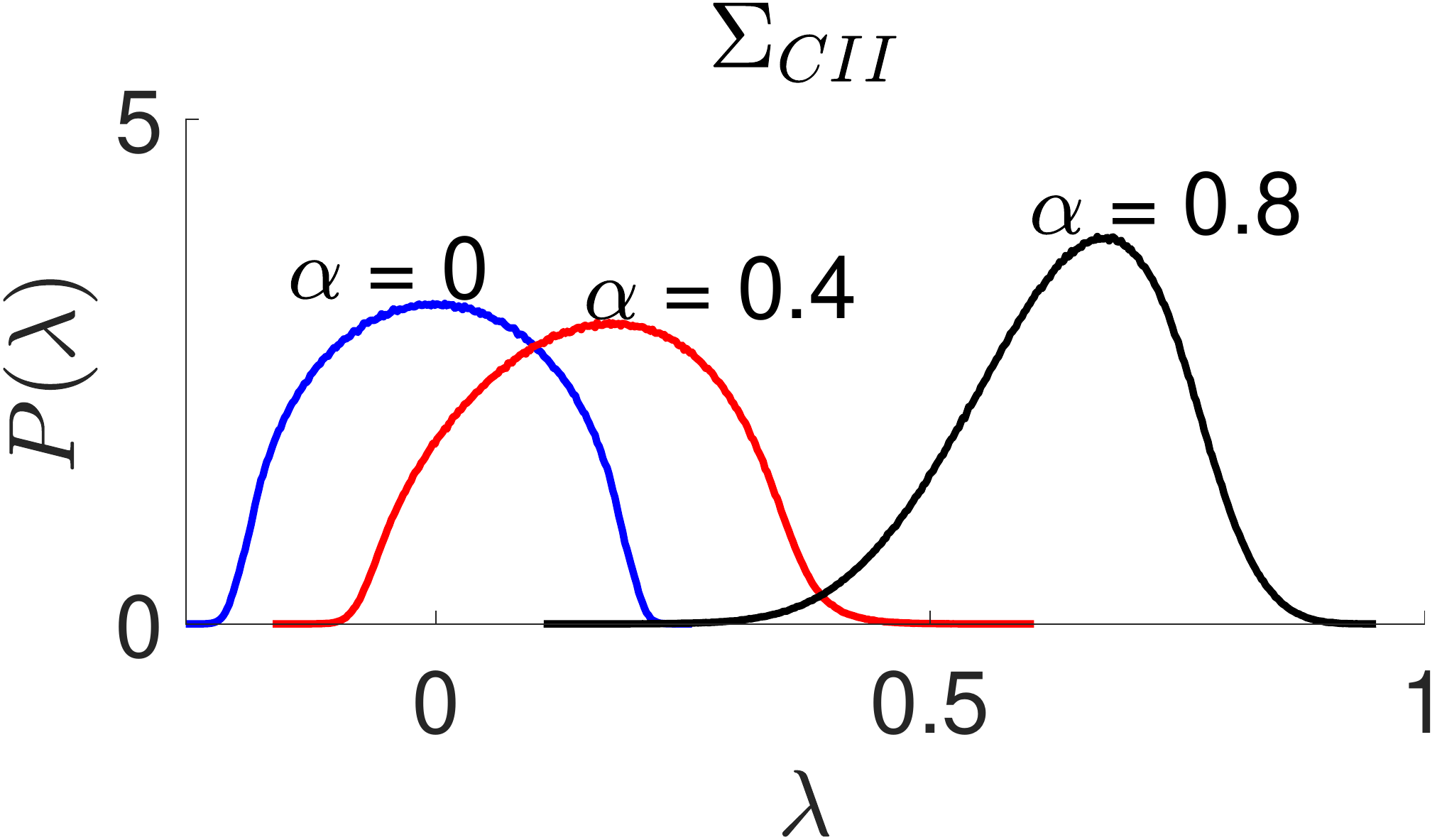}
\caption{Distribution of eigenvalues of $100\times 100$ random matrices from $\Sigma_{BDI}$, $\Sigma_{AIII}$ and $\Sigma_{CII}$. Well described by semi-circle laws for $\alpha=0$, but new distributions appear for general $\alpha$.} \label{Fig4}
\end{figure}

\section{Algebraic Preliminaries}

Average values over compact Lie groups and symmetric spaces can be computed in terms of so-called Weingarten functions, that have attracted some attention recently \cite{collins, CS, Banica, matsumoto}. 

Some concepts and notation must be introduced. Further discussion of these topics can be found, for example, in \cite{matsumoto,macdonald}

We denote by $\lambda\vdash n$ the fact that $\lambda$ partitions $n$. The permutation group on $n$ symbols is $S_n$. The character of permutation $\pi$ in the irreducible representation of $S_n$ labelled by $\lambda$ is denoted $\chi_\lambda(\pi)$, and $d_\lambda=\chi_\lambda(1)$ is the dimension of the representation. The conjugacy class in $S_n$ corresponding to cycle type $\lambda$ is denoted $C_\lambda$.

A transposition is a $2$-cycle $(i\,j)$. A permutation is even if it can be written as the product of an even number of transpositions, odd otherwise. The sign of $\sigma$, denoted $\epsilon(\sigma)$, is $1$ if $\sigma$ is even and $-1$ if $\sigma$ is odd.

We shall make use of the following specific permutations:
\begin{align} S_n\ni \pi_{U}&=(1\,2\cdots n),\\
 S_{2n}\ni \phi_{U}&=(1\,2)(3\,4)\cdots (2n-1\,2n),\\
  S_{2n}\ni \varphi_{U}&=(2\,3)(4\,5)\cdots (2n\, 1),\end{align}
  also
  \begin{align}
  S_{2n}\ni \pi_{O}&=(1\,2\cdots 2n)^2=(1\,3\,5\cdots)(2\,4\,6\cdots),\\
  S_{4n}\ni \phi_{O}&=(1\,2\,3\,4)(5\,6\,7\,8)\cdots, \\
  S_{4n}\ni \varphi_{O}&=(1\,2\,4n-1\,4n)(3\,4\,5\,6)\cdots,
\end{align}
and
  \begin{align}
  S_{4n}\ni \pi_{BDI}  = (2\,4\,5\,7\,)\cdots (4k-2\, 4k\,4k+1\,4k+3)\cdots 
  \end{align}

Matchings are partitions of the set $\{1,...,2n\}$ into $n$ blocks of size $2$. The trivial matching is $\mathfrak{t}=\{\{1,2\},\{3,4\},...,\{2n-1,2n\}\}$. They can be represented by permutations in two ways. First, a matching $\mathfrak{m}$ can be represented by the permutation $\pi$ if $\pi(\mathfrak{t})=\mathfrak{m}$, e.g. $(23)\{\{1,2\},\{3,4\}\}=\{\{1,3\},\{2,4\}\}$. The stabilizer of $\mathfrak{t}$ in $S_{2n}$ is the hyperoctahedral group, $H_n$, which has $n!2^n$ elements. We may consider as equivalent the permutations that lead to the same matching, and the set of such equivalence classes is denoted $\M_n=S_{2n}/H_n$. Second, a matching can be turned into a fixed-point-free involution in the obvious way, e.g. $\mathfrak{m}=\{\{1,2\},\{3,4\}\}\to f(\mathfrak{m})=(12)(34)$. We write $f(\sigma)$ for the fixed-point-free involution associated with the matching $\sigma(\mathfrak{t})$.

Suppose the numbers $\{1,...,2n\}$ label the vertices of a graph. Given a permutation $\sigma$, connect two of these vertices if they belong to the same block in $\mathfrak{t}$ or in $\sigma(\mathfrak{t})$. The weakly decreasing list containing half the sizes of the connected components of this graph is called the coset type of $\sigma$, denoted $[\sigma]$, and it partitions $n$.

Given two strings of $n$ symbols, $\vec{i}$ and $\vec{j}$, the function $\delta_\sigma[\vec{i},\vec{j}]$, with $\sigma\in S_n$ takes value $1$ if $\vec{j}$ is obtained from $\vec{i}$ by the permutation $\sigma$, $\vec{j}=\sigma(\vec{i})$, and vanishes otherwise. On the other hand, given a string of $2n$ symbols, $\vec{i}$, the function $\Delta_\sigma[\vec{i}]$, with $\sigma\in \M_n$ takes value $1$ if $\vec{i}$ is invariant under $f(\sigma)$, and vanishes otherwise. Moreover, the function $\Delta'_\sigma[\vec{i}]$ is defined as $\Delta'_\sigma[\vec{i}]=\prod_{r=1}^n\langle i_{\sigma(2r-1)}|i_{\sigma(2r)}\rangle$, where 
\be \langle i|j\rangle=\begin{cases}1, &\text{if $1\le i\le N$, $j=i+N$,}\\-1, &\text{if $1\le j\le N$, $i=j+N$},\\0, &\text{otherwise}.
\end{cases}\ee

Finally, Jack polynomials, $J_\lambda^{\gamma}(x)$, are defined as 
\begin{align}
J^{2}_\lambda(x)&=\sum_{\mu\vdash n}2^{n-\ell(\mu)}|C_\mu|\omega_\lambda(\mu)p_\mu(x),\\
J^{1}_\lambda(x)&=\frac{1}{d_\lambda}\sum_{\mu\vdash n}|C_\mu|\chi_\lambda(\mu)p_\mu(x),
\end{align}
where $p_\mu(x)$ are the usual power sum symmetric polynomials. The function 
\be \omega_\lambda(\tau)=\frac{2^nn!}{(2n)!}\sum_{\xi\in H_{n}}\chi_{2\lambda}(\tau\xi)\ee is the zonal spherical function of the pair ($S_{2n},H_n$) and depends only on the coset type of $\tau$. The Jack polynomial $J_\lambda^{1/2}(x)$ is also useful, but more cumbersome to define (see \cite{matsumoto}).

The functions $J^{2}_\lambda(x)$ are also known as zonal polynomials; the functions $J^{1}_\lambda(x)$ are closely related to Schur functions. Their value when all arguments are equal to $1$ is specially important,
\be J_\lambda^{\gamma}(1^N)=\gamma^{n}\prod_{i=1}^{\ell(\lambda)}\frac{\Gamma(\lambda_i+(N-i+1)/\gamma)}{\Gamma((N-i+1)/\gamma)}\ee and is a generalization of the raising factorial.

\section{Lie Groups}

The ensembles $\Sigma_O$, $\Sigma_U$ and $\Sigma_S$, associated with the classical compact Lie groups, contain non-symmetric matrices with complex spectrum. In this Section we compute the first few values of $m_n^{G}=\langle \rt M^n\rangle_{G}$ and also $s_n^{G}=\langle \rt(MM^T)^n\rangle_{G}$ for these ensembles. The results we find are consistent with universality. On the other hand, we discuss how their calculation for general $n$ is related to some difficult enumerative problems involving permutation groups.

Let us mention that a quarter-circle distribution, $\rho_Q(x)=\frac{4}{\pi X^2}\sqrt{X^2-x^2}$, has even moments given by $\int_0^X \rho_Q(x)x^{2n}dx=\frac{C_n}{4^n}X^{2n}$, where $C_n=\frac{1}{n+1}\binom{2n}{n}$ are the Catalan numbers.

\subsection{Unitary group}

For the unitary group, we have \be\label{IU} \left\langle \prod_{k=1}^n U_{i_kj_k}U^*_{q_kp_k}\right\rangle_{\U(N)} = \sum_{\sigma,\tau \in S_n} \Wg^{U}_N(\sigma^{-1}\tau)\delta_\tau[\vec{q},\vec{i}]\delta_\sigma[\vec{p},\vec{j}].\ee The coefficient $\Wg^{U}_N$, a function on the permutation group $S_n$, is the Weingarten function of $\U(N)$. It is given by
\be\label{WU} \Wg^{U}_N(\pi)=\frac{1}{n!}\sum_{\lambda\vdash n}\frac{d_\lambda}{J_\lambda^{1}(1^N)}\chi_\lambda(\pi).\ee

We want the average value of ${\rm Tr}M^n$, which can be written as \be\sum_{\vec{i}} M_{i_1i_2}M_{i_2i_3}\cdots M_{i_ni_1}=\sum_{\vec{i}} |U_{i_1i_2}|^2|U_{i_2i_3}|^2\cdots |U_{i_ni_1}|^2.\ee
This is in the form of Eq.(\ref{IU}), provided we take $\vec{q}=\vec{i}$ and $\vec{p}=\vec{j}=\pi_U(\vec{i})=(i_2,\cdots,i_n,i_1)$, with $\pi_U$ given in Section 4.  

We therefore arrive at
\be\langle {\rm Tr}M^n\rangle_{\Sigma_U}=\sum_{\vec{i}}\sum_{\sigma,\tau \in S_n} \Wg^{U}_N(\sigma^{-1}\tau)\delta_\tau[\vec{i},\vec{i}]\delta_\sigma[\pi_U(\vec{i}),\pi_U(\vec{i})].\ee The quantity \be \sum_{\vec{i}}\delta_\tau[\vec{i},\vec{i}]\delta_\sigma[\pi_U(\vec{i}),\pi_U(\vec{i})] \ee counts how many strings $\vec{i}$ are simultaneously invariant under the permutations $\tau$ and $\pi_U^{-1}\sigma\pi_U$. If we denote by $\langle a,b\rangle$ the group generated by $a$ and $b$, and by $\Omega\langle a,b\rangle$ the number of orbits of this group when acting on the set $\{1,...,n\}$, then the above quantity equals $N^{\Omega\langle \tau,\pi_U^{-1}\sigma\pi_U\rangle}$ and we end up with 
\be\left\langle {\rm Tr}M^n\right\rangle_{\Sigma_U}=\sum_{\sigma,\tau \in S_n} \Wg^{U}_N(\sigma^{-1}\tau)N^{\Omega\langle \tau,\pi_U^{-1}\sigma\pi_U\rangle}.\ee 

Since the Weingarten function $\Wg^{U}_N$ depends only on the cycle type of its argument, we could also write \be\label{mnU}\left\langle {\rm Tr}M^n\right\rangle_{\Sigma_U}=\sum_{\lambda\vdash n}\sum_{m=1}^nF^U_n(m,\lambda) \Wg^{U}_N(\lambda)N^{m},\ee where \be F^U_n(m,\lambda)=\#\{(\sigma,\tau), \pi_U^{-1}\sigma^{-1}\pi_U\tau\in C_\lambda,\Omega\langle\tau,\sigma\rangle=m\}\ee is the number of pairs $(\sigma,\tau)$ which generate a group with $m$ orbits and such that $\pi_U^{-1}\sigma^{-1}\pi_U\tau$ has cycle type $\lambda$.

The simplest term in Eq.(\ref{mnU}) comes from $\lambda=1^n$ and $m=n$. This combination arises only for $\sigma=\tau=1$, so $F^U_n(n,1^n)=1$. Since $\Wg^{U}_N(1^n)=1$, we get $\left\langle {\rm Tr}M^n\right\rangle_{\Sigma_U}=1+O(N^{-1}),$ reflecting the contribution of the Perron-Frobenius eigenvalue.

Higher-order contributions can be obtained by solving the combinatorial problem on a computer. This leads to the following tables for the functions $F^U_n(m,\lambda)$:
\be 
F_2^U= \begin{pmatrix}1&2\\ \noalign{\medskip}1&0\end{pmatrix}, \quad  F^U_3= \begin{pmatrix} 5&12&9\\ \noalign{\medskip}0&6&3
\\ \noalign{\medskip}1&0&0\end{pmatrix},\quad F^U_4=  \begin{pmatrix} 16&112&50&144&104\\ \noalign{\medskip}7&
20&20&44&40\\ \noalign{\medskip}0&12&2&4&0\\ \noalign{\medskip}1&0&0&0
&0\end{pmatrix}. 
\ee
The sum of all entries is $n!^2$ is each case. The sum of entries in the $\lambda$ column is the number of pairs $(\sigma',\tau)$ such that $\sigma'\tau\in C_\lambda$.

Using such tables, we can find for the reduced traces $m_n^{U}=\left\langle \rt M^n\right\rangle_{\Sigma_U}$:
\begin{align} m_{2}^{U}&=\frac{1}{N+1}\sim \frac{1}{N},\\ m_{3}^{U}&=\frac{2}{(N+1)(N+2)}\sim \frac{2}{N^2},\\
m_{4}^{U}&=\frac{N^2+12N+6}{N(N+1)(N+2)(N+3)}\sim \frac{1}{N^2},\\
 m_{5}^{U}&=\frac{34}{(N+1)(N+2)(N+3)(N+4)}\sim \frac{34}{N^4}
\end{align}

For the singular values, we have  \be\tr(MM^T)^n=\sum_{\vec{i},\vec{j}} M_{i_1j_1}M_{i_2j_1}M_{i_2j_2}M_{i_3j_2}\cdots M_{i_nj_n}M_{i_1j_n}.\ee Writing this in terms of the unitary matrices, we have \be\left\langle \tr(MM^T)^n\right\rangle_{\Sigma_U}=\sum_{\vec{i},\vec{j}}\sum_{\sigma,\tau \in S_{2n}} \Wg^{U}_N(\sigma^{-1}\tau)\delta_\tau[\vec{i},\vec{i}]\delta_\sigma[\vec{j},\vec{j}],\ee with the strings $\vec{i}$ and $\vec{j}$ being of the form $\vec{i}=(i_1,i_2,i_2,\cdots,i_n,i_n,i_1)$ and $\vec{j}=(j_1,j_1,j_2,j_2,\cdots,j_n,j_n)$.

We see that $\vec{i}$ must be simultaneously invariant under the actions of $\tau$ and of the permutation $\varphi_U=(2\,3)(4\,5)\cdots (2n\,1)$. Therefore, the quantity $\sum_{\vec{i}}\delta_\tau[\vec{i},\vec{i}]$ is given by $N^{\Omega\langle \tau,\varphi_U\rangle}$. Likewise, we have $\sum_{\vec{j}}\delta_\sigma[\vec{j},\vec{j}]=N^{\Omega\langle \sigma,\phi_U\rangle}$ with $\phi_U=(1\,2)(3\,4)\cdots (2n-1\,2n)$. 

Therefore, 
\be\left\langle \tr(MM^T)^n\right\rangle_{\Sigma_U}=\sum_{\sigma,\tau \in S_{2n}} \Wg^{U}_N(\sigma^{-1}\tau)N^{\Omega\langle \sigma,\phi_U\rangle+\Omega\langle \tau,\varphi_U\rangle},\ee or 
\be\label{singU}\left\langle \tr(MM^T)^n\right\rangle_{\Sigma_U}=\sum_{\lambda\vdash n}\sum_{k,m=1}^nG^U_n(m,k,\lambda) \Wg^{U}_N(\lambda)N^{m+k},\ee where \be G^U_n(m,k,\lambda)=\#\{(\sigma,\tau), \sigma^{-1}\tau\in C_\lambda,\Omega\langle \sigma,\phi_U\rangle=m,\Omega\langle \tau,\varphi_U\rangle=k\}\ee is the number of pairs $(\sigma,\tau)$ such that $\sigma^{-1}\tau$ has cycle type $\lambda$ and the groups $\langle \sigma,\phi_U\rangle$ and $\langle \tau,\varphi_U\rangle$ have $m$ and $k$ orbits, respectively.

Solving this combinatorial problem in the computer, we get for the reduced traces $s_n^{U}=\left\langle \rt(MM^T)^n\right\rangle$:
\begin{align} s_1^{U}&=\frac{N-1}{N+1}\sim 1- \frac{2}{N},\\
s_2^{U}&=\frac{2(N-1)(N+4)}{(N+3)(N+2)(N+1)}\sim \frac{2}{N},\\
s_3^{U}&=\frac{5N^4+60N^3+217N^2-46N-256}{(N+5)(N+4)(N+3)(N+2)(N+1)^2}\sim \frac{5}{N^2}.
\end{align}

The numbers $1,2,5$, which appear in the above leading-order numerators, are the first Catalan numbers, a result that is consistent with the distribution of singular values being a quarter-circle with radius $2/\sqrt{N}$.

\subsection{Orthogonal group}

For the orthogonal group, we have \be\label{IO} \left\langle \prod_{k=1}^{2n} U_{i_kj_k}\right\rangle_{\OO(N)} = \sum_{\sigma,\tau \in \M_n} \Wg^{O}_N(\sigma^{-1}\tau)\Delta_\tau[\vec{i}]\Delta_\sigma[\vec{j}].\ee The coefficient $\Wg^{O}_N$, the Weingarten function of $\OO(N)$, is given by
\be\label{WO} \Wg^{O}_N(\pi)=\frac{2^nn!}{(2n)!}\sum_{\lambda\vdash n}\frac{d_{2\lambda}}{J_\lambda^{2}(1^N)}\omega_\lambda(\pi).\ee

Writing \be \tr M^n=\sum_{\vec{i}} U_{i_1i_2}^2U_{i_2i_3}^2\cdots U_{i_ni_1}^2,\ee
we arrive at \be\langle {\rm Tr}M^n\rangle_{\Sigma_O}=\sum_{\vec{i}}\sum_{\sigma,\tau \in \M_{n}} \Wg^{O}_N(\sigma^{-1}\tau)\Delta_\tau[\vec{i}]\Delta_\sigma[\pi_O(\vec{i})],\ee where now the permutation $\pi_O$ is the square of the cycle, $\pi_O=(1\,2\,\cdots \,2n)^2$ and the string $\vec{i}$ is of the form $\vec{i}=(i_1,i_1,i_2,i_2,...,i_n,i_n)$. This last condition can be implemented by imposing that $\vec{i}$ is invariant under $\phi_U=f(1)$, where $f(\sigma)$ is, as discussed, the fixed-point-free involution associated with the matching $\sigma(\mathfrak{t})$.

The quantity \be \sum_{\vec{i}}\Delta_\tau[\vec{i}]\Delta_\sigma[\pi_O(\vec{i})]=N^{\Omega\langle f(\tau),f(\pi_O\sigma),\phi_U\rangle} \ee is given in terms of the number of orbits of the group generated by the fixed-point-free involutions associated to the permutations $\tau$, $\pi_O\sigma$ and 1. Therefore, \be\langle {\rm Tr}M^n\rangle_{\Sigma_O}=\sum_{\lambda\vdash n}\sum_{m=1}^n F_n^O(m,\lambda)W_N^O(\lambda)N^m,\ee where $F^O_n(m,\lambda)$ is the number of pairs $(\sigma,\tau)$ such that $\langle f(\tau),f(\pi_O\sigma),\phi_U\rangle$ has $m$ orbits and $\sigma^{-1}\tau$ has coset type $\lambda$.

Solving this combinatorial problem on a computer leads to the following tables for the functions $F^O_n(m,\lambda)$:
\be 
F_2^O= \begin{pmatrix}2&6\\ \noalign{\medskip}1&0\end{pmatrix}, \quad  F^O_3= \begin{pmatrix} 14&78&108\\ \noalign{\medskip}0&12&12
\\ \noalign{\medskip}1&0&0\end{pmatrix},\ee and \be F^O_4=  \begin{pmatrix} 88&1136&1112&3072&4576\\ \noalign{\medskip}16&
100&140&272&464\\ \noalign{\medskip}0&24&8&16&0\\ \noalign{\medskip}1&0&0&0
&0\end{pmatrix}. 
\ee
The sum of all entries is $(2n-1)!!^2$ in each case. The sum of entries in the $\lambda$ column is the number of pairs $(\sigma,\tau)$ such that $\sigma^{-1}\tau$ has coset type $\lambda$.

Explicit computations give
\begin{align} m_{2}^{O}&=\frac{2}{N+2}\sim \frac{2}{N},\\ m_{3}^{O}&=\frac{8}{(N+2)(N+4)}\sim \frac{8}{N^2},\\
m_{4}^{O}&=\frac{4(N^2+23N+36)}{(N+1)(N+2)(N+4)(N+6)}\sim \frac{4}{N^2},\\ 
m_{5}^{O}&=\frac{16(29N+24)}{(N+1)(N+2)(N+4)(N+6)(N+8)}\sim \frac{464}{N^4}.\end{align}

For the singular values, we have \be\left\langle \tr(MM^T)^n\right\rangle_{\Sigma_O}=\sum_{\vec{i},\vec{j}}\sum_{\sigma,\tau \in \M_{2n}} \Wg^{O}_N(\sigma^{-1}\tau)\Delta_\tau[\vec{i}]\Delta_\sigma[\vec{j}],\ee where $\vec{i}$ has length $4n$ and is of the form \be \vec{i}=(i_1,i_1,i_2,i_2,i_2,i_2,\cdots,i_n,i_n,i_n,i_n,i_1,i_1)\ee while $\vec{j}$ is of the form $ \vec{j}=(j_1,j_1,j_1,j_1,\cdots,j_n,j_n,j_n,j_n)$. So $\vec{i}$ is invariant under $\varphi_O$ and $\vec{j}$ is invariant under $\phi_O$, defined in Section 4. The quantities \be \sum_{\vec{i}}\Delta_\tau[\vec{i}]=N^{\Omega\langle f(\tau),\varphi_O\rangle} , \quad \sum_{\vec{j}}\Delta_\sigma[\vec{j}]=N^{\Omega\langle f(\sigma),\phi_O\rangle} \ee are given in terms of the number of orbits of the groups generated by  $\varphi_O$ (respectively,  $\phi_O$) and the fixed-point-free involution associated with $\tau$ (respectively, $\sigma$). 

We can write 
\be\label{singO}\left\langle \tr(MM^T)^n\right\rangle_{\Sigma_O}=\sum_{\lambda,\vdash n}\sum_{k,m=1}^n G_n^O(m,k,\lambda)\Wg^{O}_N(\sigma^{-1}\tau)N^{m+k},\ee where $G_n^O(m,k,\lambda)$ is the number of pairs $(\sigma,\tau)$ such that $\sigma^{-1}\tau$ has coset type $\lambda$ and the groups $\langle f(\tau),\varphi_O\rangle$ and $\langle f(\sigma),\phi_O\rangle$ have $m$ and $k$ orbits, respectively. Explicit computations give
\begin{align} s_1^{O}&=\frac{2N-2}{N+2}\sim 2-\frac{6}{N},\\
s_2^{O}&=\frac{(4N-4)(2N^2+17N+12)}{(N+1)(N+2)(N+4)(N+6)}\sim \frac{8}{N},\end{align} where $s_n^{O}=\langle \rt (MM^T)^n\rangle_{\Sigma_O}$.

The above leading-order numerators, $2$ and $8$, are consistent with the distribution of singular values being a quarter-circle with radius $2\sqrt{2/N}$.

\subsection{Symplectic group}

For the symplectic group, we have \be\label{IS} \left\langle \prod_{k=1}^{2n} U_{i_kj_k}\right\rangle_{Sp(2N)} = \sum_{\sigma,\tau \in \M_n} \Wg^{Sp}_N(\sigma^{-1}\tau)\Delta'_\tau[\vec{i}]\Delta'_\sigma[\vec{j}].\ee The coefficient $\Wg^{Sp}_N$, the Weingarten function of $Sp(2N)$, can be found in \cite{matsumoto} and is related to the orthogonal one as $\Wg^{Sp}_N(\sigma)=(-1)^n\epsilon(\sigma)\Wg^{O}_{-2N}(\sigma)$, where $\epsilon(\sigma)$ is the sign of $\sigma$.

Writing \begin{align} \tr M^n&=\sum_{\vec{i}} U_{i_1i_2}U^D_{i_2i_1}U_{i_2i_3}U^D_{i_3i_2}\cdots U_{i_ni_1}U^D_{i_1i_n}\\
&=\sum_{\vec{i}} U_{i_1i_2}U_{i_1+N,i_2+N}U_{i_2i_3}U_{i_2+N,i_3+N}\cdots U_{i_ni_1}U_{i_n+N,i_1+N},\end{align}
where now each index is summed from $1$ to $2N$ (and addition is performed modulo $2N$), we arrive at \be \left\langle {\rm Tr}M^n\right\rangle_{\Sigma_S}=\sum_{\vec{i}}\sum_{\sigma,\tau \in \M_{n}} \Wg^{Sp}_N(\sigma^{-1}\tau)\Delta'_\tau[\vec{i}]\Delta'_\sigma[\pi_O(\vec{i})],\ee where $\pi_O=(1\,2\,\cdots \,2n)^2$ and the string $\vec{i}$ is of the form $\vec{i}=(i_1,i_1+N,i_2,i_2+N,...)$. 

The quantity $\sum_{\vec{i}}\Delta'_\tau[\vec{i}]\Delta'_\sigma[\pi_O(\vec{i})]$ is, up to sign, given by $(2N)^{\Omega\langle f(\tau),f(\pi_O\sigma),\phi_U\rangle}$ and is almost equal to the one that appeared for the orthogonal group. This sign is a function of the coset types of $\tau$ and $\pi_O\sigma$ and also depends on the sequence $\vec{i}$. This prevents us from analysing this quantity further. We have only been able to compute
\be 
m^S_2 = \frac{2}{2N+1}\sim \frac{1}{N}.
\ee

For the singular values, we have \be \left\langle \tr(MM^T)^n\right\rangle_{\Sigma_S}=\sum_{\vec{i},\vec{j}}\sum_{\sigma,\tau \in \M_{2n}} \Wg^{Sp}_N(\sigma^{-1}\tau)\Delta'_\tau[\vec{i}]\Delta'_\sigma[\vec{j}],\ee where $\vec{i}$ has length $4n$ and is of the form \be\vec{i}=(i_1,i_1+N,\cdots,i_n,i_n+N,i_n,i_n+N,i_1,i_1+N)\ee and \be\vec{j}=(j_1,j_1+N,j_1,j_1+N,\cdots,j_n,j_n+N,j_n,j_n+N).\ee We have only been able to compute
\be
s^S_1  = \frac{2N^2+N+1}{(N-1)(2N+1)}\sim 1+\frac{1}{N}.
\ee

\section{Circular Ensembles}

Unlike the ensembles $\Sigma_O$, $\Sigma_U$ and $\Sigma_S$, associated with the groups, the ensembles $\Sigma_{AI}$ and $\Sigma_{AII}$ contain symmetric matrices with real spectrum. In this Section we compute the first few values of $m_n^{G}=\langle {\rm Tr}M^n\rangle_{G}$ for these ensembles. The results we find are consistent with the observed universal semi-circle distribution of eigenvalues.

Let us mention that a semi-circle distribution, $\rho_{SC}(x)=\frac{2}{\pi X^2}\sqrt{X^2-x^2}$, has even moments given by $\int_{-X}^X \rho_{SC}(x)x^{2n}dx=\frac{C_n}{4^n}X^{2n}$, where $C_n=\frac{1}{n+1}\binom{2n}{n}$ are the Catalan numbers.

The calculation of spectral moments for general $n$ again leads to some interesting, but rather difficult, combinatorial problems involving permutations.

\subsection{Symmetric space AI}

As a representative of each coset in the symmetric space $AI=\U(N)/\OO(N)$, one can take $U=VV^T$, with $V\in \U(N)$. With this prescription, it follows \cite{matsumoto} that \be\label{IAI} \left\langle \prod_{k=1}^{n} U_{i_{2k-1}i_{2k}}U^*_{j_{2k-1}j_{2k}}\right\rangle_{AI(N)} = \sum_{\sigma \in S_{2n}} \Wg^{AI}_N(\sigma)\delta_\sigma[\vec{i},\vec{j}],\ee where the coefficient $\Wg^{AI}_N$, the Weingarten function of AI, is given in terms of the corresponding function for the orthogonal group as $\Wg^{AI}_N(\sigma)=\Wg^{O}_{N+1}(\sigma)$.

We write $M_{ij}=|U_{ij}|^2$ (note that $U$ is symmetric and so is $M$) and \be \tr M^n=\sum_{\vec{i}} U_{i_1i_2}U_{i_2i_3}\cdots U_{i_{n}i_{1}}U^*_{i_1i_2}U^*_{i_2i_3}\cdots U^*_{i_{n}i_1},\ee
we arrive at \be\left\langle {\rm Tr}M^n\right\rangle_{\Sigma_{AI}}=\sum_{\vec{i}}\sum_{\sigma\in S_{2n}} \Wg^{AI}_N(\sigma)\delta_\sigma[\vec{i},\vec{i}],\ee where $\vec{i}$ is of the form $\vec{i}=(i_1,i_2,i_2,\cdots, i_n,i_n,i_1)$, invariant under the action of $\varphi_U$.

Since the Weingarten function $\Wg^{AI}_N$ depends only on the coset type of its argument, we can also write \be\label{MAI}\left\langle \tr M^n\right\rangle_{\Sigma_{AI}}=\sum_{\lambda\vdash n}\sum_{m=1}^nF^{AI}_n(m,\lambda) \Wg^{AI}_N(\lambda)N^{m},\ee where \be F^{AI}_n(m,\lambda)=\#\{\sigma,[\sigma]=\lambda,\Omega\langle\sigma,\varphi_U\rangle=m\}\ee is the number of permutations $\sigma$ with coset type $\lambda$ and such that $\langle\sigma,\varphi_U\rangle$ has $m$ orbits.

The contribution with $\lambda=1^n$ and $m=n$ gives $\left\langle \tr M^n\right\rangle_{\Sigma_{AI}}=1+O(N^{-1}),$ reflecting the Perron-Frobenius eigenvalue.

Higher-order contributions can be obtained by solving the combinatorial problem on a computer. This leads to the following tables for the functions $F^{AI}_n(m,\lambda)$:
\begin{align}
F^{AI}_2 &= \begin{pmatrix}6&14\\ \noalign{\medskip}2&2\end{pmatrix}, \quad  F^{AI}_3= \begin{pmatrix} 38&234&320\\ \noalign{\medskip}9&51&60
\\ \noalign{\medskip}1&3&4\end{pmatrix},\end{align} and \be F^{AI}_4 =  \begin{pmatrix} 306&3800&3862&10312&15608\\ \noalign{\medskip}67&
724&689&1820&2636\\ \noalign{\medskip}10&80&54&152&184\\ \noalign{\medskip}1&4&3&4
&4\end{pmatrix}. 
\ee
The sum of all entries is $(2n)!$ is each case. The sum of entries in the $\lambda$ column is the number of permutations $\sigma$ with coset type $\lambda$.

Using such tables, we can find for the reduced traces $m_n^{AI}=\left\langle \rt M^n\right\rangle_{\Sigma_{AI}}$:
\begin{align} m_{1}^{AI}&=\frac{N-1}{N+1}\sim 1-\frac{2}{N},\\
m_{2}^{AI}&=\frac{(N-1)(N+5)}{(N+1)(N+3)}\sim 1-\frac{8}{N^2},\\
m_{3}^{AI}&=\frac{3N^2+22N-29}{(N+1)(N+3)(N+5)}\sim \frac{3}{N}-\frac{5}{N^2},\\
m_{4}^{AI}&=\frac{2(N^4+20N^3+146N^2+92N-323)}{(N+1)(N+2)(N+3)(N+5)(N+7)}\sim \frac{2}{N}+\frac{4}{N^2},\end{align}

The existence of a continuous density of eigenvalues of order $1/\sqrt{N}$ would imply that $\frac{1}{N} \left\langle \rt M^n\right\rangle\sim \frac{1}{N^{n/2}}$ and hence $N^{n/2-1}m_n^{AI}$ should have a finite limit as $N\to \infty$. We see that this is true for the even moments and we see the Catalan numbers $1$ and $2$ appear, in agreement with the semi-circle law. On the other hand, for the odd moments this limit vanishes, indicating a symmetric distribution. 

Notice however that, for finite $N$, the first moment is positive, corresponding to a small shift of the distribution towards positive values which can be seen in Figure \ref{Fig3}.

\subsection{Symmetric space AII}

As a representative of each coset in the symmetric space $AII=\U(2N)/Sp(2N)$, one can take $U=VV^DJ$, with $V\in \U(N)$ and $J=\left( \begin{array}{cc} 0_N & I_N \\ -I_N & 0_N
\end{array}  \right)$. This prescription, slightly different from the one in \cite{matsumoto}, leads to antisymmetric matrices $U$ which satisfy \be\label{IAII} \left\langle \prod_{k=1}^{n} U_{i_{2k-1}i_{2k}}U^*_{j_{2k-1}j_{2k}}\right\rangle_{AII(N)} = \sum_{\sigma \in S_{2n}} \Wg^{AII}_N(\sigma)\delta_\sigma[\vec{i},\vec{j}],\ee where the Weingarten function of AII is  given in terms of the corresponding function for the symplectic group as $\Wg^{AII}_N(\sigma)=\Wg^{S}_{N-1/2}(\sigma)=(-1)^n\epsilon(\sigma)\Wg^{O}_{1-2N}(\sigma)$.

Writing $M_{ij}=|U_{ij}|^2$ (since $U$ is antisymmetric, $M$ is symmetric) and \be \tr M^n=\sum_{\vec{i}} U_{i_1i_2}U_{i_2i_3}\cdots U_{i_{n}i_{1}}U^*_{i_1i_2}U^*_{i_2i_3}\cdots U^*_{i_{n}i_1},\ee
we arrive at \be\left\langle {\rm Tr}M^n\right\rangle_{\Sigma_{AII}}=\sum_{\vec{i}}\sum_{\sigma\in S_{2n}} \Wg^{AII}_N(\sigma)\delta_\sigma[\vec{i},\vec{i}],\ee where $\vec{i}$ is again invariant under the action of $\varphi_U$. The Weingarten function $\Wg^{AII}_N(\sigma)$ depends on the coset type of $\sigma$ and on its sign, $\epsilon(\sigma)$. Hence, we can write \be\left\langle \tr M^n\right\rangle_{\Sigma_{AII}}=(-1)^n\sum_{\lambda\vdash n}\sum_{m=1}^nF^{AII}_n(m,\lambda)\Wg^{O}_{1-2N}(\lambda)N^{m},\ee where $F^{AII}_n(m,\lambda)=F^{AII}_{n,+}(m,\lambda)-F^{AII}_{n,-}(m,\lambda)$ is now the difference between the solutions of two combinatorial problems, 
\be F_{n,\pm}^{AII}(m, \lambda)= \# \lbrace \sigma \in S_{2n} | [\sigma] = \lambda , \Omega \langle\sigma, \varphi_U\rangle=m , \epsilon(\sigma)=\pm 1 \rbrace\ee namely the number of permutations $\sigma$, positive or negative, with coset type $\lambda$ such that the group $\langle\sigma, \varphi_U\rangle$ has $m$ orbits.

The corresponding tables are
\be 
 F^{AII}_2 =\begin{pmatrix} -2 &2  \\ 2 & -2 \end{pmatrix}, \quad F^{AII}_3 =\begin{pmatrix}  -2 & 6 & -4 \\ 3 & -9 & 6 \\ -1 & 3  & -2 \end{pmatrix},\ee and \be
F^{AII}_4 =\begin{pmatrix} -14  & 72  & -42  & -88  & 72 \\ 11 & -44 & 33 & 44 & -44 \\ 2 & -24 & 6 & 40 & -24 \\ 1 & -4 & 3 & 4 & -4  \end{pmatrix}.\ee

The first reduced traces are:
\begin{align} m_{1}^{AII}&=-1,\\
m_{2}^{AII}&=1,\\
m_{3}^{AII}&=\frac{-3}{2N+1}\sim -\frac{3}{2N}+\frac{3}{(2N)^2},\\
m_{4}^{AII}&=\frac{2N+5}{(2N+1)(N+1)}\sim \frac{2}{2N}+\frac{4}{(2N)^2}.\end{align}
Similarly to what happens for $\Sigma_{AI}$, this indicates for large $N$ a continuous density of eigenvalues of order $1/\sqrt{N}$, symmetric, with even moments consistent with a semi-circle of radius $2/\sqrt{N}$. For finite $N$, however, the first moment is negative and a small shift toward negative values can indeed be noticed in Figure \ref{Fig3}. 

\section{Chiral Ensembles}

In contrast to previous cases, the chiral ensembles have two parameters, $a$ and $b$, with $a+b=N$. In the large $N$ limit, it is natural to define $\alpha=(a-b)/N$ and fix it. For all values of $\alpha$, the matrices in $\Sigma_{AIII}$, $\Sigma_{BDI}$ and $\Sigma_{CII}$ are symmetric and have a real spectrum. 

\subsection{Symmetric space AIII}

Elements in $AIII=\U(N)/\U(a)\times \U(b)$, with $a+b=N$, can be taken in the form $U=V\tilde{J}_aV^\dagger$, with $V\in \U(N)$ and $\tilde{J}_a=I_a\oplus (-I_{b})$. Since $U$ is hermitian, the matrix elements of the associated stochastic matrix can be written as $M_{ij}=|U_{ij}|^2=U_{ij}U_{ji}$.

In this ensemble we have \cite{matsumoto} \be\label{IAIII} \left\langle \prod_{k=1}^{n} U_{i_kj_k}\right\rangle_{AIII(N)} = \sum_{\sigma \in S_{n}} \Wg^{AIII}_{N,a,b}(\sigma)\delta_\sigma[\vec{i},\vec{j}],\ee where the Weingarten function $\Wg^{AIII}_{N,a,b}$ is given by:
\be \Wg^{AIII}_{N,a,b}(\sigma)=\frac{1}{(2n!)}\sum_{\lambda\vdash n}d_\lambda\frac{J_\lambda^1(1^a,(-1)^b)}{J_\lambda^1(1^{a+b})}\chi_\lambda(\sigma). \ee

Writing \be \tr M^n=\sum_{\vec{i}} U_{i_1i_2}U_{i_2i_1}U_{i_2i_3}U_{i_3i_2}\cdots U_{i_{n}i_{1}}U_{i_1i_n},\ee
we arrive at \be\left\langle {\rm Tr}M^n\right\rangle_{\Sigma_{AIII}}=\sum_{\vec{i}}\sum_{\sigma\in S_{n}} \Wg^{AIII}_{N,a,b}(\sigma)\delta_\sigma[\vec{i},\phi_U{i}],\ee where $\vec{i}=(i_1,i_2,i_2,\cdots,i_n,i_n,i_1)$ is again invariant under the action of $\varphi_U$. The Weingarten function $\Wg^{AIII}_N$ depends only on the cycle type of its argument, hence \be\left\langle \tr M^n\right\rangle_{\Sigma_{AIII}}=\sum_{\lambda\vdash 2n}\sum_{m=1}^nF^{AIII}_n(m,\lambda) \Wg^{AIII}_{N,a,b}(\lambda)N^{m},\ee where $F^{AIII}_n(m,\lambda)$ is the number of permutations $\sigma$ with cycle type $\lambda$ and such that $\langle\sigma\phi_U,\varphi_U\rangle$ has $m$ orbits.

\be F^{AIII}_1=\begin{pmatrix}1 & 1\end{pmatrix}, \quad F^{AIII}_2= \begin{pmatrix}
1 & 6 & 1 & 8 & 4 \\
0 & 0 & 2 & 0 & 2
\end{pmatrix},\ee
and
\be F^{AIII}_3=\begin{pmatrix} 
1 & 15 & 39 & 11 & 40 & 96 & 30 & 84 & 66 & 120 & 90\\
0 & 0  & 6  & 3  & 0  & 24 & 9  & 6  & 21 & 24  & 27\\
0 & 0  & 0  & 1  & 0  & 0  & 1  & 0  & 3  & 0   & 3 
\end{pmatrix}.\ee

In terms of the parameter $\alpha=(a-b)/N$, of order $1$, the first reduced traces are:
\begin{align} m_{1}^{AIII}&=\frac{N^2\alpha^2-1}{N+1}\sim \alpha^2(N-1),\\
m_{2}^{AIII}&=\frac{\alpha^4N^3+(2\alpha^2+1)N^2-(4\alpha^2-3)N-3}{(N+1)(N+3)}\sim \alpha^4N-(4\alpha^4-2\alpha^2-1).\end{align} The exact expression is too cumbersome for the next ones, but for large $N$
\begin{align}
m_{3}^{AIII}&\sim\alpha^{6}N-\alpha^2(10\alpha^4-6\alpha^2-3)+\frac{1}{N}\alpha^2(\alpha^2-1)(72\alpha^2-1)\\
m_{4}^{AIII}&\sim\alpha^{8}N-\alpha^4(19\alpha^4-12\alpha^2-6)+\frac{2}{N}(\alpha^2-1)(124\alpha^6-16\alpha^4-5\alpha^2-1).
\end{align} We conjecture that $m_{n}^{AIII}\sim \alpha^{2n}N$.

For large $N$, the average eigenvalue is simply $\alpha^2$, and it seems that $\lim_{N\to\infty}\frac{1}{N}m_{n}^{AIII}=\alpha^{2n}$, suggesting the convergence in distribution to a $\delta$-function. Therefore, it makes sense to compute the shifted moments $\mu_n^{AIII}=\langle\rt (M-\alpha^2)^n\rangle$. The first one is 
\be
\mu_1^{AIII}= \frac{\alpha^2-1}{N+1}, 
\ee
while the next ones are asymptotic to
\begin{align}
\mu_2^{AIII}&\sim -\left(\alpha^2-1\right)\left(3\alpha^2+1\right) + \frac{\left( \alpha^2-1\right)\left(11\alpha^2 -1 \right)}{N}\\
\mu_3^{AIII}&\sim \frac{\left(\alpha^2-1\right) \left(36\alpha^4-4\alpha^2\right)}{N} \\
\mu_4^{AIII}&\sim \frac{2\left( \alpha^2-1\right)^2 \left(17\alpha^4+6\alpha^2+1\right)}{N}
\end{align}

These results suggest that the eigenvalues scale as $1/\sqrt{N}$ around the average. At the symmetry point $\alpha=0$ the semi-circle law seems valid, based on the fact that, at least for $n=1$ and $n=2$, we have $\frac{1}{N}\mu_{2n}^{AIII}\sim C_n/N^n$, where $C_n$ are the Catalan numbers.

\subsection{Symmetric space BDI}

Elements in $BDI=\OO(N)/\OO(a)\times \OO(b)$, can be taken in the form $U=V\tilde{J}_aV^T$, with $V\in \OO(N)$, in which case they are real and symmetric. Then, $M_{ij}=|U_{ij}|^2=U_{ij}^2$.

In this ensemble we have \cite{matsumoto} \be\label{IBDI} \left\langle \prod_{k=1}^n U_{i_{2k-1}i_{2k}}\right\rangle_{BDI(N)} = \sum_{\sigma \in \M_{n}} \Wg^{BDI}_{N,a,b}(\sigma)\Delta_\sigma[\vec{i}],\ee where the Weingarten function $\Wg^{BDI}_{N,a,b}$ is given by:
\be \Wg^{BDI}_{N,a,b}(\sigma)=\frac{2^nn!}{(2n!)}\sum_{\lambda\vdash n}d_{2\lambda}\frac{J_\lambda^2(1^a,(-1)^b)}{J_\lambda^2(1^{a+b})}\omega_\lambda(\sigma). \ee

Writing \be \tr M^n=\sum_{\vec{i}} U_{i_1i_2}^2U_{i_2i_3}^2\cdots U_{i_{n}i_{1}}^2,\ee
we arrive at \be\left\langle {\rm Tr}M^n\right\rangle_{\Sigma_{BDI}}=\sum_{\vec{i}}\sum_{\sigma\in \M_{2n}} \Wg^{BDI}_{N,a,b}(\sigma)\Delta_\sigma[\vec{i}],\ee where $\vec{i}=(i_1,i_2,i_1,i_2,i_2,i_3,i_2,i_3,\cdots)$, which is invariant under $\pi_{BDI}$.

Since $\Wg^{BDI}_N$ depends only on the coset type of its argument, \be\left\langle \tr M^n\right\rangle_{\Sigma_{BDI}}=\sum_{\lambda\vdash 2n}\sum_{m=1}^nF^{BDI}_n(m,\lambda) \Wg^{BDI}_{N,a,b}(\lambda)N^{m},\ee where $F^{BDI}_n(m,\lambda)$ is the number of matchings $\sigma$ with coset type $\lambda$ and such that $\langle f(\sigma),\pi_{BDI}\rangle$ has $m$ orbits. The first few solutions of this combinatorial problem are given by 
$$F^{BDI}_1= \left( \begin{array}{cc}
1 & 2 \\
0 & 0
\end{array} \right), \quad F^{BDI}_2= \left( \begin{array}{ccccc}
1 & 12 & 9 & 32 & 42 \\
0 & 0 & 3 & 0 & 6 \\
0 & 0 & 0 & 0 & 0 \\
0 & 0 & 0 & 0 & 0 
\end{array} \right)$$

In terms of the parameter $\alpha=(a-b)/N$, the first reduced traces are:
\begin{align} 
m_{1}^{BDI}&= \frac{N^2\alpha^2+N-2}{N+2} \sim \alpha^2\left(N-2\right)+1+\frac{4\left(\alpha^2-1\right)}{N},\\
m_{2}^{BDI}& \sim \left(N-9\right)\alpha^4+6\alpha^2+2+\frac{\left(\alpha^2-1\right)\left(61\alpha^2+3\right)}{N}  .
\end{align} 
The average eigenvalue is again $\alpha^2$. The first shifted moments are 
\begin{align}
\mu_1^{BDI}&= -\frac{\left(\alpha^2-1\right)\left(N-2\right)}{N+2} \thicksim -\left(\alpha^2-1\right) + \frac{4\left(\alpha^2-1\right)}{N}, \\
\mu_2^{BDI}&\thicksim -\left(\alpha^2 -1\right)\left(6\alpha^2+2\right) - \frac{\left(\alpha^2-1\right)\left(\alpha^2+15\right)}{N}.
\end{align}

\subsection{Symmetric space CII}

Elements in $CII=Sp(2N)/Sp(2a)\times Sp(2b)$, can be taken in the form $U=V\tilde{K}_aV^D$, with $V\in Sp(2N)$, in which case $U$ is self-dual. Then, $M_{ij}=|U_{ij}|^2=U_{ij}U_{j+N i+N}$.

In this ensemble we have \cite{matsumoto} \be\label{ICII} \left\langle \prod_{k=1}^n U_{i_{2k-1}+Ni_{2k}}\right\rangle_{CII(N)} = \sum_{\sigma \in \M_{n}} \Wg^{CII}_{N,a,b}(\sigma)\Delta'_\sigma[\vec{i}],\ee where the Weingarten function $\Wg^{CII}_{N,a,b}$ can be found in \cite{matsumoto}.

Writing \be \tr M^n=\sum_{\vec{i}} U_{i_1i_2} U_{i_2+Ni_1+N}U_{i_2i_3}U_{i_3+Ni_2+N}\cdots U_{i_{n}i_{1}}U_{i_{1}+Ni_{n}+N},\ee
we arrive at \be\left\langle {\rm Tr}M^n\right\rangle_{\Sigma_{CII}}=\sum_{\vec{i}}\sum_{\sigma\in \M_{n}} \Wg^{CII}_{N,a,b}(\sigma)\Delta'_\sigma[\vec{i}].\ee 

Form the above expression we can compute the first reduced moment \be m^{CII}_1= \frac {4\,{N}^{3}{\alpha}^{2}-4\,{N}^{2}-N+1}{ \left( N-1 \right) 
 \left(2\,N+1 \right) } \sim \alpha^2(2N)+{\alpha}^{2}-2+\frac{3}{2}\frac {({\alpha}^{2}-1)}{N}
 \ee
and the first shifted moment 
\be \mu^{CII} = \frac { \left( \alpha^2 -1 \right)   \left( 4\,{ N}^{2}+N-1 \right) }{ \left( N-1 \right)  \left( 2\,N+1 \right) }
 \thicksim  2\left({\alpha}^{2}-1\right)+\frac{3}{2}\frac {({\alpha}^{2}-1)}{N}.
\ee

\section{The combinatorial problems}

For convenience, we collect here the combinatorial problems that have appeared in our calculations. Solutions for the simplest cases have been presented in the text. Notation was set in Section 4.1.

The calculation of $m_n=\langle {\rm Tr}M^n\rangle$ requires
\begin{itemize}

\item for $\Sigma_U$, the number of pairs of permutations $(\sigma,\tau)$ such that $\pi_U^{-1}\sigma^{-1}\pi_U\tau$ has cycle type $\lambda$ and $\langle \sigma,\tau\rangle$ has $m$ orbits. 

\item for $\Sigma_O$, the number of pairs of matchings $(\sigma,\tau)$ such that $\sigma^{-1}\tau$ has coset type $\lambda$ and $\langle f(\tau),f(\pi_O\sigma),\phi_U\rangle$ has $m$ orbits.

\item for $\Sigma_{AI}$ and $\Sigma_{AII}$, the number of permutations $\sigma$ with coset type $\lambda$ such that $\langle \sigma,\varphi_U\rangle$ has $m$ orbits.

\item for $\Sigma_{AIII}$, the number of permutations $\sigma$ with cycle type $\lambda$ such that $\langle \sigma\phi_U,\varphi_U\rangle$ has $m$ orbits.

\item for $\Sigma_{BDI}$, the number of matchings $\sigma$ with coset type $\lambda$ such that $\langle f(\sigma), \pi_{BDI}\rangle$ has $m$ orbits.

\end{itemize}

The calculation of $s_n=\langle {\rm Tr}(MM^T)^n\rangle$ requires
\begin{itemize}

\item for $\Sigma_U$, the number of pairs of permutations $(\sigma,\tau)$ such that $\sigma^{-1}\tau$ has cycle type $\lambda$, $\langle \sigma,\phi_U\rangle$ has $m$ orbits and $\langle \tau,\varphi_U\rangle$ has $k$ orbits.

\item for $\Sigma_O$, the number of pairs of matchings $(\sigma,\tau)$ such that $\sigma^{-1}\tau$ has coset type $\lambda$, $\langle \sigma,\phi_O\rangle$ has $m$ orbits and $\langle \tau,\varphi_O\rangle$ has $k$ orbits.

\end{itemize}

\section{Some large $N$ asymptotics and cancellations}

The large $N$ asymptotics of our results involves the large $N$ behavior of Weingarten  functions. This problem has indeed attracted attention. 

Matsumoto and Novak have shown \cite{matnovak} that \be W_N^U(\lambda)=(-1)^{n-\ell(\lambda)}\sum_{q=0}^\infty c_{\lambda,q}^U N^{-n-q},\ee where $c_{\lambda,q}^U$ is the number of length $q$ monotone factorizations of a permutation with cycle type $\lambda$. Other expansions for  $W_N^U(\lambda)$ were considered by Collins \cite{collins}, Berkolaiko and Irving \cite{irving} and Novaes \cite{expan}.

Matsumoto showed \cite{mato} that \be W_N^O(\lambda)=\sum_{q=0}^\infty(-1)^q c_{\lambda,q}^O N^{-n-q},\ee where $c_{\lambda,q}^O$ is the number of some length $q$ monotone factorizations of a permutation with coset type $\lambda$.  

Berkolaiko and Kuipers showed \cite{kuipers} that \be W_N^{AI}(\lambda)=\sum_{q=0}^\infty(-1)^q c_{\lambda,q}^{AI} N^{-n-q},\ee where $c_{\lambda,q}^{AI}$ is the number of some length $q$ palindromic monotone factorizations. Another expansion for  $W_N^{AI}(\lambda)$ was considered by Novaes \cite{expan}.

To review all these factorization problems would lead us too far away. We refer to the original papers for details. Interestingly, when we consider the large $N$ expansion for $\Sigma_U$, $\Sigma_O$ and $\Sigma_{AI}$, these problems become intermingled with the combinatorial problems  listed in the previous Section.

From (\ref{singU}), we have 
\be \frac{1}{N}\langle \tr (MM^T)^n\rangle_{\Sigma_U}=\sum_{j=1}^\infty T_{n,j}^U N^{n-j},\ee
where the coefficients are given by
\be\label{AU}  T_{n,j}^U=\sum_{\lambda}(-1)^{n-\ell(\lambda)}\sum_{k,m=1}^nG_n^U(m,k,\lambda)c_{\lambda,m+k+j-2n-1}^U. \ee
The conjectured quarter-circle distribution implies $\frac{1}{N}\langle \tr (MM^T)^n\rangle_{\Sigma_U}\sim 1+C_nN^{-n}$. This means not only that $T_{n,2n}^U=C_n$ is the Catalan number, but also that cancellations in (\ref{AU}) lead to $T_{n,j}^U=0$ for all $j<2n$.

From (\ref{singO}), we have 
\be \frac{1}{N}\langle \tr (MM^T)^n\rangle_{\Sigma_O}=\sum_{j=1}^\infty T_{n,j}^O N^{n-j},\ee
where the coefficients are given by
\be\label{AO}  T_{n,j}^O=\sum_{\lambda}\sum_{k,m=1}^n(-1)^{m+k+j-1}G_n^O(m,k,\lambda)c_{\lambda,m+k+j-2n-1}^O. \ee
The conjectured quarter-circle distribution in this case implies $\frac{1}{N}\langle \tr (MM^T)^n\rangle_{\Sigma_O}\sim 1+2^nC_nN^{-n}$. This means not only that $T_{n,2n}^O=2^nC_n$, but also that cancellations in (\ref{AO}) lead to $T_{n,j}^O=0$ for all $j<2n$.

From (\ref{MAI}), we have 
\be \frac{1}{N}\langle \tr M^{2n}\rangle_{\Sigma_{AI}}=\sum_{j=1}^\infty T_{n,j}^{AI} N^{n-j},\ee
where the coefficients are given by
\be\label{AAI}  T_{n,j}^{AI}=\sum_{\lambda}\sum_{m=1}^{2n}(-1)^{m+j+n-1}F_n^{AI}(m,\lambda)c_{\lambda,m+j-3n-1}^{AI}. \ee
The conjectured semi-circle distribution implies $\frac{1}{N}\langle \tr M^{2n}\rangle_{\Sigma_{AI}}\sim 1+C_nN^{-n}$. This means not only that $T_{n,2n}^{AI}=C_n$, but also that cancellations in (\ref{AAI}) lead to $T_{n,j}^{AI}=0$ for all $j<2n$.

In all cases, we have the appearance of Catalan numbers in $T_{n,2n}$ and the vanishing of $T_{n,j}$ for all $j<2n$. These seem to be rather strong results with several implications for the combinatorics of permutations. Perhaps they deserve further study. 

\section{Conclusions}

We have associated ensembles of stochastic matrices to classical compact Lie groups and some symmetric spaces. For the Lie groups, the corresponding stochastic matrices are not symmetric and have complex spectrum, while they are symmetric and have real spectrum for matrices obtained from the symmetric spaces. 

Numerically, we observed that the macroscopic statistical properties of the reduced spectrum of our random stochastic matrices conform to the universal results which are expected (those of the real Ginibre ensemble for the Lie groups and those of the Gaussian Orthogonal Ensemble for the symmetric spaces). However, universality is only a conjecture in this case, since our matrices do not have independent elements.

We approached the calculation of $\langle {\rm Tr}M^n\rangle$ and $\langle {\rm Tr}(MM^T)^n\rangle$ via Weingarten functions. Explicit calculation is possible for the first few cases and agrees with the conjectured universality. A thorough understanding of these quantities might open the way to exact results on spectral statistics. However, their calculation is related to combinatorial problems involving permutations.

These problems are difficult, because they require knowledge of the number of orbits of a group given its generators, which is already a non-trivial question, coupled with conditions on the cycle type or coset type of the generators.

The large $N$ regime, moreover, brings into play the asymptotics of Weingarten functions, which are related to factorizations of permutations. The conjectured universality for spectral statistics is then translated into some subtle relations between these two kinds of combinatorial problems. 

M.N. was supported by grants 303634/2015-4 and 400906/2016-3 from Conselho Nacional de Desenvolvimento Cient\'ifico e Tecnol\'ogico (CNPq). L.H.O. was supported by a fellowship from Coordena\c{c}\~ao de Aperfei\c{c}oamento de Pessoal de N\'ivel Superior.


\begin{thebibliography}{99}

\bibitem{horvat} Horvat M 2009 J. Stat. Mech. P07005  
\bibitem{chafai} Chafa\"i D 2010 J. Multivariate Analysis {\bf 101} 555 
\bibitem{bordenave} Bordenave C, Caputo P, Chafa\"i D 2012 Probab. Th. Related Fields {\bf 152} 751
\bibitem{inocentini} Innocentini GCP, Novaes M 2018 J. Stat. Mech. 103202

\bibitem{diac1} Chatterjee S, Diaconis P, Sly A arXiv:1010.6136

\bibitem{bi} Cappellini V, Sommers H-J, Bruzda W and  \.{Z}yczkowski K 2009 J. Phys. A: Math. Theor. {\bf 42} 365209

\bibitem{uzy} Kottos T and Smilansky U 1997 Phys. Rev. Lett. {\bf 79} 4794

\bibitem{tanner1} Tanner G 2000 J. Phys. A: Math. Gen. {\bf 33} 3567

\bibitem{tanner2} Tanner G 2001 J. Phys. A: Math. Gen. {\bf 34} 8485

\bibitem{greg} Berkolaiko G 2001 J. Phys. A: Math. Gen. {\bf 34} L319

\bibitem{zycz} \.{Z}yczkowski K, Ku\'s M, S\l omczy\'nski W and Sommers H-J 2003 J. Phys. A: Math. Gen. {\bf 36} 3425 

\bibitem{circ1} Edelman A 1997 J. Multivariate Anal. {\bf 60}, 203

\bibitem{circ2} Kanzieper E and Akemann G 2005 Phys. Rev. Lett. {\bf 95}, 230201

\bibitem{edelman} Edelman A, Kostlan E and Shub M 1994 J. Amer. Math. Soc. 7, 247 

\bibitem{wigner0} Pastur L 1973 Teoret. Mat. Fiz. {\bf 10}, 102

\bibitem{wigner1} Erd\"os L, Ramirez J, Schlein B, Tao T, Vu V and Yau H-T 2010 Math. Res. Lett. {\bf 17}, 667

\bibitem{wigner2} Tao T and Vu V arXiv:0906.0510.

\bibitem{collins} Collins B 2003 Int. Math. Res. Not. {\bf 17} 953 

\bibitem{CS} Collins B and \'Sniady P 2006 Comm. Math. Phys. {\bf 264} 773

\bibitem{Banica} Banica T, Collins B and Schlenker J-M 2011 On polynomial integrals over the orthogonal group. J. Combinat. Theory A {\bf 118} 78

\bibitem{matsumoto} Matsumoto S 2013 Random Matrices: Th. Appl. {\bf 2} 1350001

\bibitem{mezzadri} Mezzadri F 2007 Notices Am. Math. Soc. {\bf 54}, 592

\bibitem{macdonald} I.G. Macdonald, {\it Symmetric Functions and Hall Polynomials}, 2nd ed. (Oxford University Press, Oxford, 1995).

\bibitem{matnovak} Matsumoto S and Novak J 2013 Int. Math. Res. Not. 2 362

\bibitem{irving} Berkolaiko G and Irving J 2016 J. Combin. Theory A 140 1–37

\bibitem{mato} Matsumoto S 2011 Ramanujan J. 26 69

\bibitem{kuipers} Berkolaiko G and Kuipers J 2013 J. Math. Phys. 54 112103

\bibitem{expan} Novaes M 2017 J. Phys. A: Math. Theor. 50 075201

\end{thebibliography}
\end{document}